\documentclass[twocolumn]{aastex62}
\usepackage{float}
\usepackage{afterpage}

\shorttitle{Draft}
\shortauthors{Sarkar et al.}

\begin{document}

\title{\textbf{Evolution of Coronal Cavity from Quiescent to Eruptive Phase in Association with Coronal Mass Ejection}}

\correspondingauthor{Ranadeep Sarkar}
\email{ranadeep@prl.res.in}
 \author[0000-0002-0786-7307]{Ranadeep Sarkar}
 \affil{Udaipur Solar Observatory, Physical Research Laboratory,
 Badi Road, Udaipur 313001, India} \\

 \author{Nandita Srivastava}
\affiliation{Udaipur Solar Observatory, Physical Research Laboratory,
Badi Road, Udaipur 313001, India}
 

 \author{Marilena Mierla}
 \affiliation{Solar-Terrestrial Center of Excellence - SIDC, Royal Observatory of Belgium, 1180 Brussels, Belgium}
 
\affiliation{Institute of Geodynamics of the Romanian Academy, Bucharest, Romania}

 \author{Matthew J West}

 \author{Elke D'Huys}
 \affiliation{Solar-Terrestrial Center of Excellence - SIDC, Royal Observatory of Belgium, 1180 Brussels, Belgium}




\begin{abstract}

We present the evolution of a coronal cavity encompassing its quiescent and eruptive phases in the lower corona. Using the multi-vantage point observations from the SDO/AIA, STEREO SECCHI/EUVI and PROBA2/SWAP EUV imagers, we capture the sequence of quasi-static equilibria of the quiescent cavity which exhibited a slow rise and expansion phase during its passage on the solar disc from 2010 May 30 to 2010 June 13. By comparing the decay-index profiles of the cavity system during the different stages of its quiescent and pre-eruptive phases we find that the decay-index value at the cavity centroid height can be used as a good indicator to predict the cavity eruption in the context of torus instability. Combining the observations of SWAP and LASCO C2/C3 we show the evolution of the EUV cavity into the white-light cavity as a three part structure of the associated CME observed to erupt on 2010 June 13. By applying successive geometrical fits to the cavity morphology we find that the cavity exhibited non self-similar expansion in the lower corona, below 2.2 $\pm$ 0.2 R$_S$, which points to the spatial scale for the radius of source surface where the coronal magnetic field lines are believed to become radial. Furthermore, the kinematic study of the erupting cavity captures both the ``impulsive" and ``residual" phases of acceleration along with a strong deflection of the cavity at 1.3 R$_S$. We also discuss the role of driving forces behind the dynamics of the morphological and kinematic evolution of the cavity.

\end{abstract}




\section{Introduction}\label{intro}
Coronal mass ejections (CMEs) are one of the most energetic phenomena that occur in the solar atmosphere, which in turn may have severe space-weather impacts on Earth \citep{gosling,siscoe,daglis}. Therefore, understanding the physical processes behind the initiation and triggering mechanism of CMEs has become the top priority in space-science research.

Observing the development of coronal cavities in lower coronal regions can provide intriguing clues to the genesis of CMEs. Cavities are believed to be the density depleted cross section of the magnetic flux ropes, where the magnetic field strength has attained a much higher value compared to the background corona \citep{Low1995,Laurel}. They are usually observed as dark ellipses or partial ellipses at the limb in white light \citep{1970W}, soft X-ray (SXR) \citep{Vaiana1973} and extreme ultraviolet (EUV) \citep{schmahl1979} observations. 

These cavities may remain stable for days or even weeks \citep{gibson2006}, and they manifest themselves as one of the three part structure of CMEs if they erupt \citep{Fisher1981,Illing,H1999,Y2002,Maricic2004,Sterling2004,Vr2004,gibson2006,Vourlidas,Howard}. Therefore, the morphological and magnetic properties of coronal cavities hold clues to the state of the pre-eruptive equilibria and the triggering mechanism behind the CME initiation.

Earlier studies indicate that the initiation and main acceleration phase of CMEs mostly occurs
below 2 - 3 Solar Radii (R$_S$) \citep{Mac1983,chen2003,joshi2011}. Therefore, by the time CMEs have evolved into the field-of-view (FOV) of LASCO C2 which observes from 2 R$_S$, the initiation and impulsive acceleration phase of most of the CMEs cannot be captured. 

The PROBA2 \citep{Santandrea} ``Sun Watcher using APS and Image Processing" (SWAP) EUV imager (\citealt{SWAP1}, \citealt{SWAP2}) provides an extended (54 arc-minute) FOV of the lower solar atmosphere, which gives the opportunity to capture the evolution of erupting cavities upto 1.7 R$_S$ (upto 1.9 R$_S$ along the diagonal direction of the images) around the 17.4 nm wavelength bandpass. Apart from that, using the off-pointing ability of PROBA2, the SWAP FOV can be shifted in any direction in order to track coronal features of interest, upto more than 2 R$_S$. Therefore, it fills the observational gap between 1 to 2 R$_S$. When combined with LASCO C2/C3 white-light observations \citep{Brueckner} this enables us to study the complete evolution of erupting cavities. The large FOV of SWAP observations has been used in several studies of large coronal structures. Such as, capturing the different phases of prominence eruptions (e.g. \citealt{Marilena}) and the evolution of a large-scale coronal pseudo-streamer in association with cavity system \citep{Guennou}. SWAP observations also have been used in conjunction with the ground-based Mauna Loa Solar Observatory (MLSO) Mark-IV K-coronameter (Mk4: \citealt{Elmore}) to study the initiation phase of a two stage eruptive event \citep{Byrne2014}.  

Statistical studies have been made in order to distinguish the morphological characteristics of eruptive and non-eruptive cavities \citep{Forland2013}. Using the AIA 193 \AA\ passband observations, \citet{Forland2013} studied the morphological structure of 129  EUV cavities and found that cavities with a teardrop-shaped morphology are more likely to erupt. On the other hand, the partly or completely eruption of white-light cavities which formed CMEs have been studied using the combined observations of MK4 and LASCO coronagraphs \citep{gibson2006,Liu}. Comparing the MK4 observations of pre-eruptive white-light cavities with those observed a few days before the eruption, \citet{gibson2006} found that the cavities show an increase in height in the days leading up to an eruption. However, due to the absence of multiple line-of-sight observations, they concluded that it is hard to determine whether those changes in cavity height are due to the true evolution of a rising cavity or the appearance of a higher portion of the three-dimensional cavity along the line-of-sight. \citet{gibson2010} proposed an observationally constrained three-dimensional (3D) cavity model which reproduced the observed cavity morphology reasonably well, as seen from the different view-points of the two STEREO spacecraft. However, the temporal evolution of roughly the same portion of the 3D cavity morphology, starting from its long lived quiescent phase to the eruptive phase in association with CMEs, has not been previously reported. 

In this work, we present a comprehensive study of a coronal cavity which exhibited almost a two-weeks long quiescent phase on the visible solar disk and finally erupted from the north-west solar limb on 2010 June 13. Using the SDO/AIA observations, \citet{reignier2011} studied the spatial relationship of this coronal cavity together with its associated prominence structure during the pre-eruptive phase. They reported the presence of magnetic curvature forces that balance the gravitational force in order to hold the cold and dense plasma of the prominence material underlying the cavity. However, using the multi-vantage point observations from SDO/AIA, STEREO A/B and PROBA2/SWAP EUV imager during the long lived quiescent phase and combining the FOV of SWAP, LASCO C2/C3 during the eruptive phase, we study the complete evolution of the cavity with an objective to address the following key questions regarding the genesis of CMEs:
 
(i) Does the morphological evolution of the quiescent cavity hold clues to the underlying magnetostatic equilibria of the cavity system? 

(ii) What determines the initiation height of CMEs? 

(iii) What are the conditions that can lead to a cavity eruption?

(iv) How do EUV cavities seen in the lower corona evolve into the white light cavities 
seen during CME eruptions?

(v) How do the magnetic forces drive the ``impulsive" and ``residual" acceleration phases of the CME? 

(vi) Do the CMEs undergo significant deflection in the lower corona?

(vii) Do the CMEs exhibit self-similar expansion in the lower corona? If 
not, then what is the critical height above which its nature of expansion is self-similar?
 
To answer these questions, we have organized this paper as follows. In section \ref{sec:first} we present the observations of a quiescent cavity. We discuss the dynamics of that cavity during the eruptive phase in section \ref{sec:eruptive}. In section \ref{sec:results} we present the results based on our analysis. Finally, we summarize and discuss the implications of these results in section 5.  
 
\section{Observations of the coronal cavity during quiescent phase} \label{sec:first}

\begin{figure}[!h]
\centering
\includegraphics[width=0.5\textwidth]{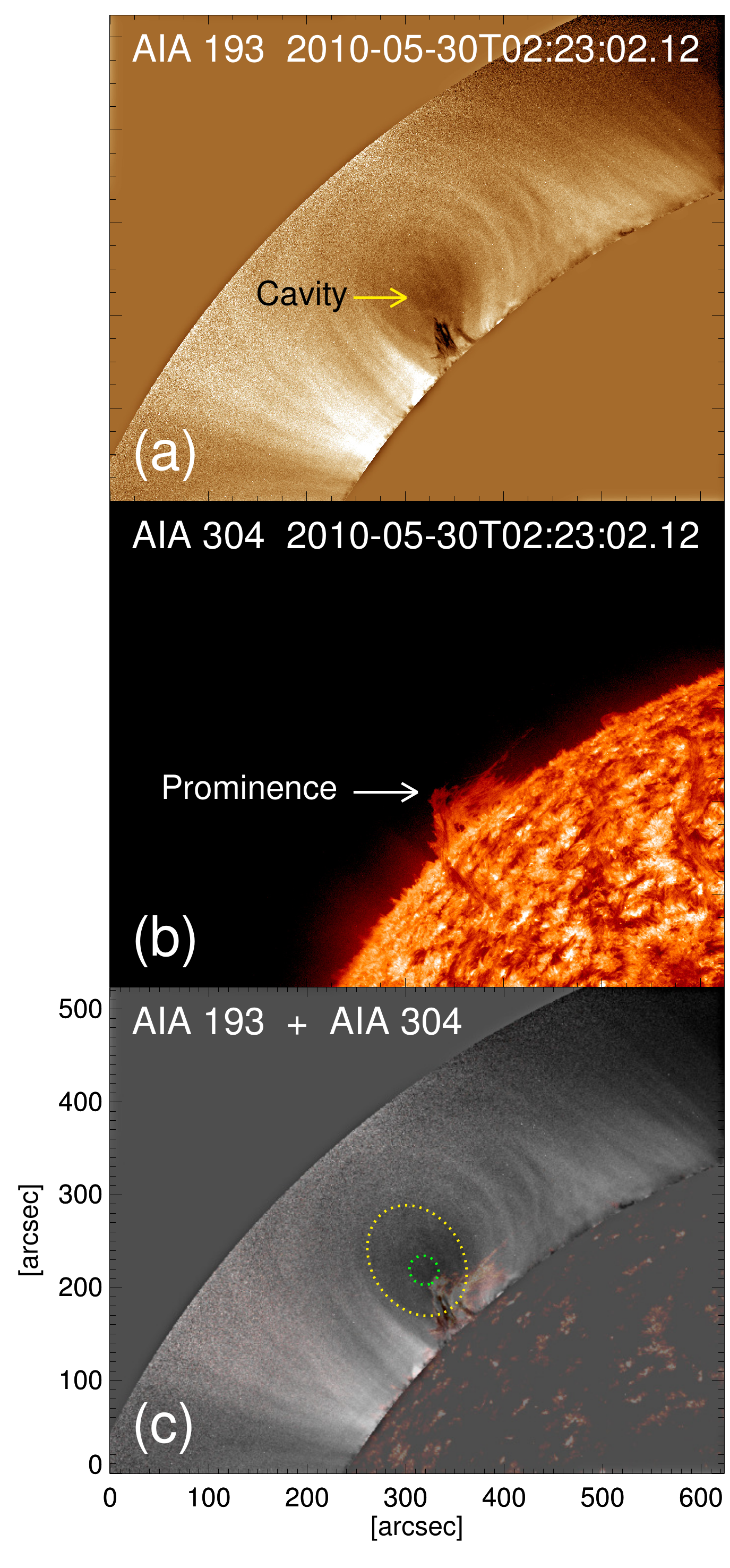}
\caption{Observation of the coronal cavity in the AIA 193 \AA\ channel (a). The associated prominence structure as seen in the AIA 304 \AA\ channel (b). The superimposed images of panels (a) and (b) are shown in panel (c). The green dotted line denotes the outer boundary of the true cavity. The yellow dotted line depicts the approximate outer boundary of the flux-rope. \label{fig:image_east_limb}}
\end{figure}

\begin{figure}[!t]
\centering
\includegraphics[width=0.45\textwidth]{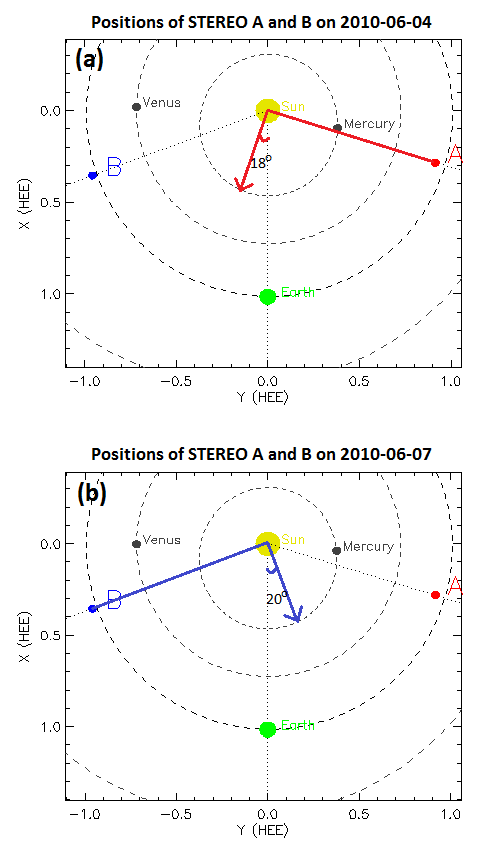}
\caption{Positions of STEREO A and B. The red and blue arrows drawn on (a) and (b) denote the direction of the plane-of-sky as viewed by EUVI on 2010 June 4 by STEREO A and on 2010 June 7 by STEREO B. \label{fig:st_combine_new}}
\end{figure}

The evolution of the coronal cavity in the lower corona during the different stages in its quiescent phase was well captured by the  SDO/AIA (\citealt{Lemen}), STEREO SECCHI/EUVI (\citealt{STEREO1}, \citealt{STEREO2}) and PROBA2/SWAP EUV imagers. The cavity was first observed on the north-east solar limb on 2010 May 30 in association with a northern polar crown filament (Figure \ref{fig:image_east_limb}). Over the period from 2010 May 30 to 2010 June 13, the cavity rotated across the solar disk and remained in a quiescent phase before its eruption on 2010 June 13  at around 6:30 UT close to the north-west solar limb.

\begin{figure*}[h]
\centering
\includegraphics[width=\textwidth]{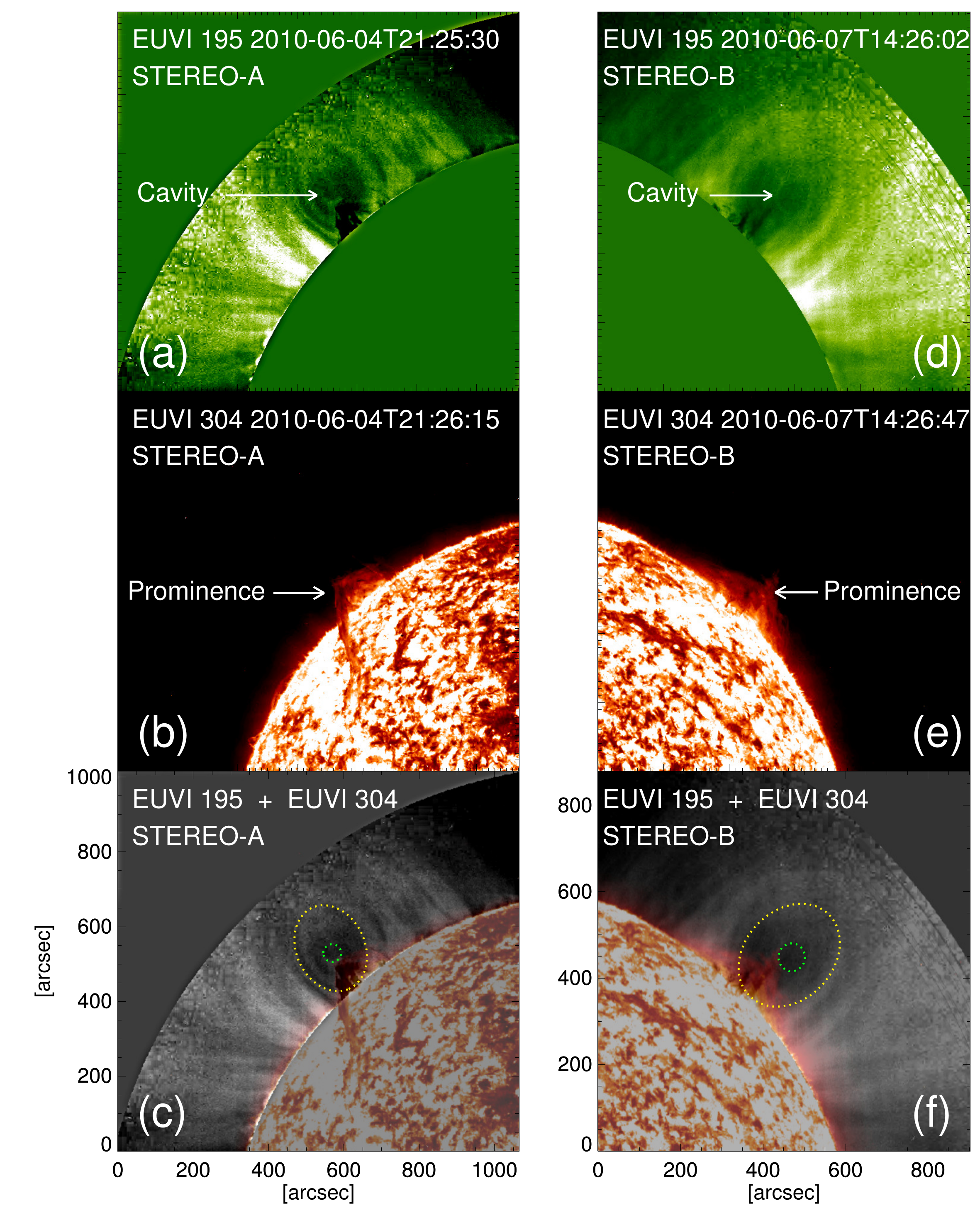}
\caption{Observations of the coronal cavity as seen in the 195 \AA\ bandpass of EUVI on STEREO A and B (panels (a) and (d) respectively), the associated prominence structure in 304 \AA\ bandpass (panels (b) and (e)) and a superposition of the 195 \AA\ and 304 \AA\ bandpasses (panels (c) and (f)). The green and yellow dotted lines are the same as those described for figure \ref{fig:image_east_limb}. \label{fig:image_stereo_limb}}
\end{figure*}

\begin{figure*}[!t]
\includegraphics[width=\textwidth]{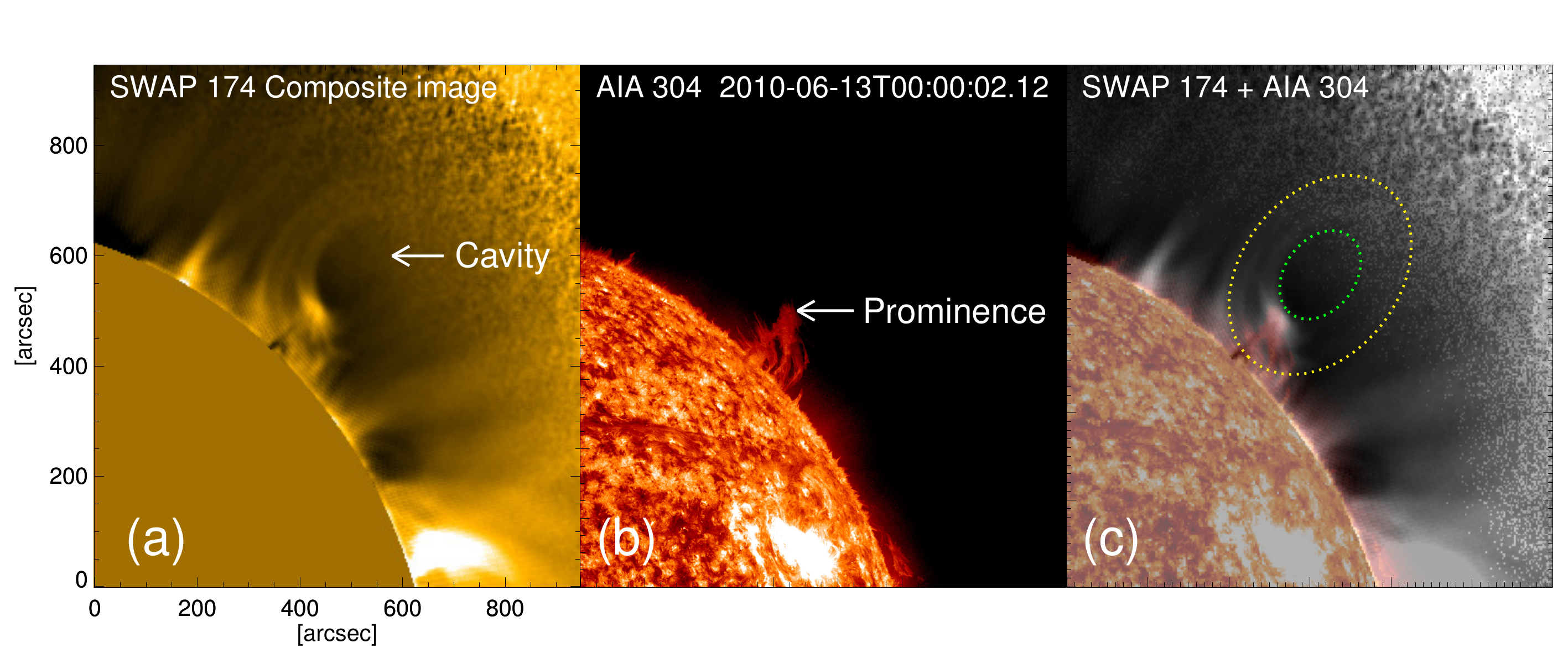}
\caption{Observations of the coronal cavity as viewed in SWAP composite images (a) and the associated prominence structure as seen in the AIA 304 \AA\ channel (b). The superimposed images of cavity morphology and the prominence structure as depicted in panels (a) and (b) are shown (c). In panel (c), the background image in gray scale represents the cavity morphology as depicted in panel (a) and the foreground image in AIA 304 \AA\ color scale represents the prominence structure as shown in panel (b). The green and yellow dotted lines are the same as those described for figure \ref{fig:image_east_limb}. \label{fig:swap_cavity}}
\end{figure*}

In order to increase visibility of coronal structures in the EUV images, we have used the normalized radial graded filter (NRGF) (\citealt{NRGF}), which removes the radial gradient from coronal images. Furthermore, we have applied a high pass filter to the EUV images using the standard  \href{https://lancesimms.com/programs/IDL/lib/unsharp_mask.pro}{unsharp\textunderscore mask.pro} routine to enhance the cavity morphology. In order to investigate the spatial association between the prominence material and the dark cavity, we have superimposed the image of prominence structure observed in the 304 \AA\ channel on top of the dark cavity as observed in the 193 \AA, 195 \AA\  and 174 \AA\ channels of AIA, EUVI and SWAP respectively (see Figures \ref{fig:image_east_limb}, \ref{fig:image_stereo_limb} and \ref{fig:swap_cavity}). 

Figure \ref{fig:image_east_limb} shows the appearance of the dark cavity over the east solar limb as seen in the AIA 193 \AA\ channel on 2010 May 30 within the FOV between 1.0 to 1.3 R$_\textrm{s}$. The polar crown prominence structure associated with this cavity system can be seen in AIA 304 \AA\ channel observations (see panel (b) in Figure \ref{fig:image_east_limb}). In order to distinguish the true cavity from the outside region of dipped field-lines which carry the prominence material, we have marked the cavity morphology, as observed in the AIA 193 \AA\ image, with the green dotted boundary drawn on the superimposed co-temporal images taken in  AIA 193 and 304 \AA\ channels (see panel (c) in Figure \ref{fig:image_east_limb}). Here we refer to the ``true'' cavity as the central non-dipped part of the magnetic flux-rope following the classification described in \citet{Gibson2015}. The combined images of the prominence structure and the cavity morphology, seen in Figure \ref{fig:image_east_limb}, indicates that the cavity is located exactly on the top of the prominence boundary. The yellow dotted elliptical boundary outside the dark cavity approximately encloses the outer boundary of the flux-rope. Outside the yellow dotted boundary the bright arched like structures resemble the overlying magnetic field lines.     

During its passage across the solar disk, the polar crown filament was well observed from multiple view points of SDO, PROBA2 and the twin spacecraft STEREO A and B. As it passed close to the central meridian, as seen from the perspective of Earth, the associated cavity morphology became visible in the plane-of-sky (POS) EUVI observations from the STEREO satellites. At that time STEREO A and B were positioned approximately 70$^\circ$ from the Sun-Earth line (see Figure \ref{fig:st_combine_new}). On 2010 June 4 the cavity appeared in the POS, from the perspective of STEREO-A (Figure \ref{fig:image_stereo_limb}), which was approximately 18$^\circ$ east of the Sun-Earth line (see Figure \ref{fig:st_combine_new}). Between 2010 June 4 to 2010 June 7, the cavity crossed the solar disk center as viewed by Earth and rotated about 20$^\circ$ further from the Sun-Earth line towards the west. At around 14:00 UT on 2010 June 7 it became visible in POS EUVI observations from STEREO-B. Figure \ref{fig:image_stereo_limb} shows the appearance of the dark cavity and the associated prominence structure as observed from the two perspectives of STEREO A and B in the 195 \AA\ (top panels) and 304 \AA\ (middle panels) bandpasses respectively. The green and yellow dotted boundaries, highlighting the true cavity and flux-rope outer boundary respectively, drawn on the superimposed images (bottom panels of Figure \ref{fig:image_stereo_limb}) are the same as those described for figure \ref{fig:image_east_limb}.




As the filament rotated towards the west solar limb, the associated cavity started to appear on the north-west solar limb as viewed from the perspective of Earth. The cavity can be observed from 2010 June 11, and is clearest at the end of 2010 June 12. Figure \ref{fig:swap_cavity} shows the cavity morphology as observed in the SWAP 174 \AA\ (left panel), the associated prominence structure in AIA 304 \AA\ (middle panel) and a combination of both bandpasses (right panel). As one of the major goals of this paper is to track the complete evolution of the eruptive cavity, we take the advantage of SWAP's large FOV, the ability to track the cavity out to $\approx$ 1.7 R$_S$ (compared to the AIA FOV which is limited to 1.3 R$_S$), and focus on SWAP observations throughout the cavity evolution from the quiescent to eruptive phase on the north-west solar limb. In order to increase the signal-to-noise in the far field of the SWAP images, the 1.6 minute cadence images between 00:00 UT and 01:35 UT on 2010 June 13 have been processed using a median stacking technique to capture the cavity morphology before eruption. Finally, the cavity erupted at around 06:30 UT from the north-west solar limb as viewed by Earth and evolved into a CME as observed in LASCO C2/C3 images.

In order to associate the morphological evolution of the EUV cavity with that of the three part structure of the associated CME seen in white light observations, it is important to understand whether the cross-sectional profiles of the cavity, as observed in EUV and white-light images, resemble the same morphological structure or not. Comparing the EUV and white-light observations of an erupting loop system that formed a CME, \citet{Byrne2014} found an inconsistency between the evolutionary profiles of the erupting structure as seen in EUV and white-light images. It is important to note that the white-light observations, which capture the Thomson scattered light from the free electrons of the solar corona, are dependent on the electron density and are more sensitive to features near the POS \citep{Vourlidas2006,Deforest_2012,Inhester}. Whereas, the EUV observations are primarily sensitive to both the temperature and density of the plasma \citep{AIA_diagnostics,DelZanna2018}, and are less preferentially sensitive to features based on their location with respect to the POS, as it is the case for white-light coronagraph images. Therefore, any coronal feature, such as the erupting loop system studied in \citet{Byrne2014}, which lies away from the POS will appear as different morphological structures in EUV and white-light observations.

Coronal cavities are extended tunnel like structures which are mostly associated with the polar crown filaments \citep{Gibson2015}. When these large structures line up along the line-of-sight, it appears like a dark croissant-like feature in the POS observations due to the density depletion in comparison to the surrounding corona. In particular, any portion of the cavity is best seen when it lies on the POS \citep{gibson2010}. Therefore, the line-of-sight integrated Thomson scattered brightness, as obtained by the white-light coronagraphs, and the line-of-sight integrated EUV emission as obtained by SWAP or AIA in 174 \AA\ and 193 \AA\ passbands which are also sensitive to the electron density due to the presence of Fe IX and Fe XII emission lines \citep{AIA_diagnostics}, should have an identical morphology for any observed coronal cavity. In order to validate this we have compared the SWAP EUV observations of the cavity morphology with the white-light observations as obtained from the Mk4 coronagraph, which observes between 1.1 to 2.8 R$_S$.

\begin{deluxetable*}{cccccc}[!t]
\tablecaption{Cavity morphology throughout the quiescent phase\label{tab:table1}}
\tablehead{
\colhead{Observation time} & \colhead{Observing instrument} &\colhead{longitude of the cavity} & \colhead{Cavity centroid} & \colhead{Cavity} & \colhead{Cavity}\\
\colhead{yyyy/mm/dd hh:mm:ss} & \colhead{ } &\colhead{in Stonyhurst /  Carrington} & \colhead{height [R$_S$]} & \colhead{morphology} & \colhead{diameter [R$_S$]}\\
\colhead{} & \colhead{} & \colhead{heliographic coordinates} & \colhead{} & \colhead{} & \colhead{} 
}

\startdata
2010/05/30 02:32:02 & SDO/AIA & 90$^\circ$ E / 136$^\circ$ &  1.10  & $\approx$ Circular & 0.016$\pm$ 0.002\\
2010/06/04 21:25:30 & STEREO-A/SECCHI EUVI & 18$^\circ$ E / 132$^\circ$ &  1.13  & $\approx$ Circular & 0.039$\pm$ 0.008\\
2010/06/07 14:26:02 & STEREO-B/SECCHI EUVI & 20$^\circ$ W / 134$^\circ$ &  1.17  & $\approx$ Circular & 0.054$\pm$ 0.008\\
2010/06/13 04:00:00 & PROBA2/SWAP & 90$^\circ$ W / 132$^\circ$ &  1.23 $\pm$ 0.02 & Elliptical         & 0.09$\pm$ 0.02\\
\enddata

\end{deluxetable*}

\begin{figure*}[!ht]
\centering
\includegraphics[width=.9\textwidth]{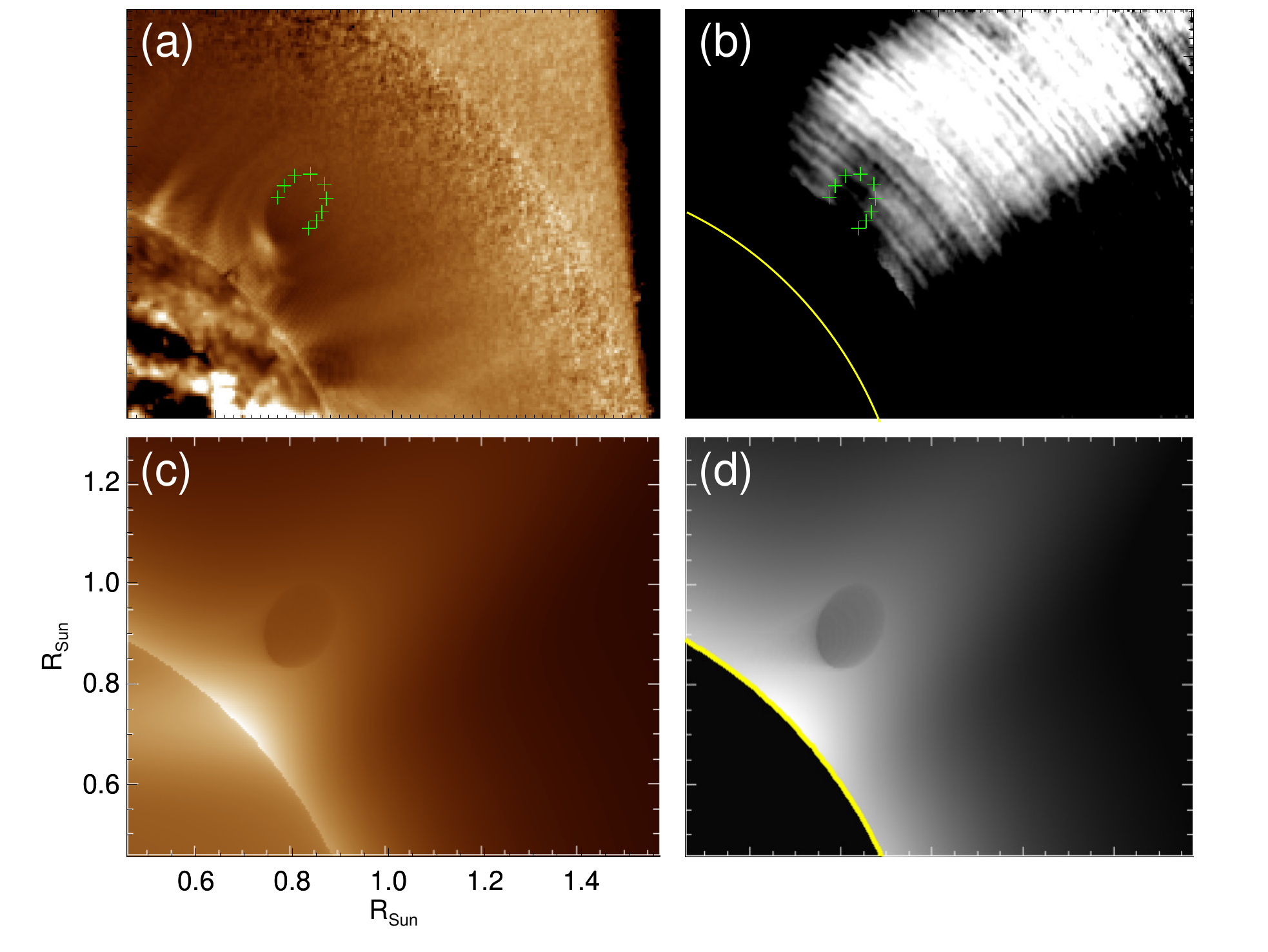}
\caption{Cavity morphology as seen in SWAP composite images stacked over the period between 00:00 UT to 01:35 UT on 2010 June 13 (a). The daily averaged polarized brightness as imaged by the groundbased coronagraph MK4 on 2010 June 13 (b). The green cross-marks drawn on panel (b) approximately indicate the outer-boundary of the white-light cavity embedded in a coronal streamer. The same green cross-marks shown in panel (b) have been drawn on panel (a). FORWARD-modeled (line-of-sight integrated) EUV emission in 174 \AA\ passband (c) and the white-light polarized brightness (d) using the model density and temperature of the cavity embedded in a coronal streamer.\label{fig:synthetic_cavity}}
\end{figure*}

Panel (b) in Figure \ref{fig:synthetic_cavity} shows the cavity morphology as seen in the daily averaged polarized brightness observations obtained from the ground-based Mk4 coronagraph on 2010 June 13. The green cross-marks drawn on panel (b) approximately indicate the outer-boundary of the white-light cavity embedded in a coronal streamer. The same green cross-marks have been over-plotted on the SWAP EUV image in panel (a), which clearly shows that the EUV and white-light cavity morphologies are the same. Furthermore, using an observationally constrained 3D cavity model \citep{gibson2010} we have FORWARD \citep{Gibson2016} modeled the cavity morphology in both the EUV emission lines and the Thomson scattered polarized brightness. The line-of-sight integrated EUV emission in 174 \AA\ passband (panel (c)) and the Thomson scattered polarized brightness (panel (d)) has been obtained using the model density and temperature of the coronal cavity, embedded in a coronal streamer \citep{Gibson1999}. The similarity between the cavity morphologies seen in the synthesized EUV and white-light images reveals that both the EUV and white-light cavities possess a fundamentally identical morphology, colocated in a large 3D structure.

\section{Evolution of the coronal cavity during eruptive phase} \label{sec:eruptive}
\subsection{Morphological Evolution of the cavity}

\begin{figure}[!ht]
\centering
\includegraphics[width=0.45\textwidth]{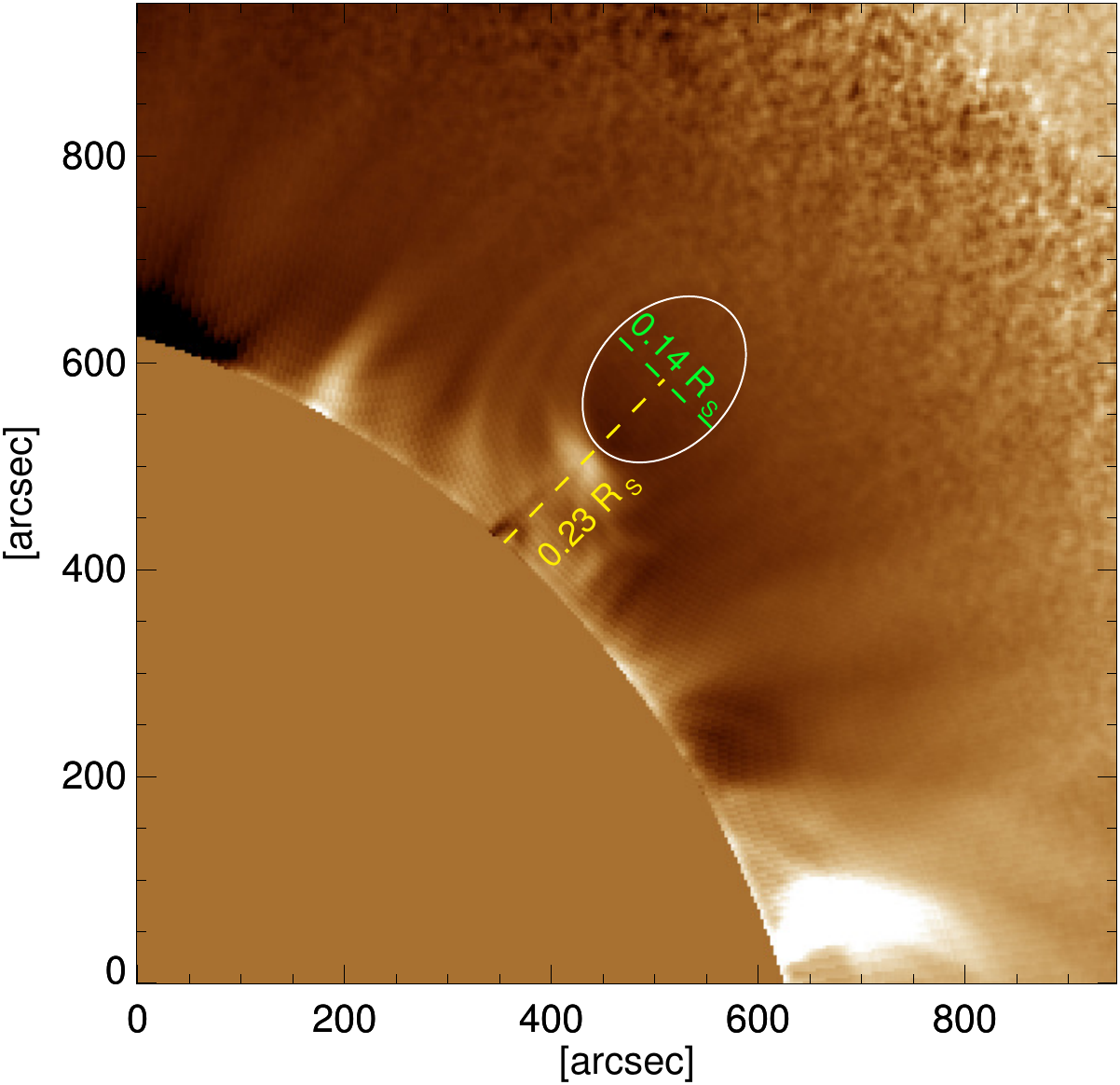}
\caption{Cavity morphology as seen in SWAP composite images stacked over the period between 00:00 UT to 01:35 UT on 2010 June 13. The cross sectional dimension of the cavity and the distance to the cavity centroid from the solar surface are indicated.\label{fig:cavity_ellipse_fitting}}
\end{figure}

\begin{figure}[!h]
\centering
\includegraphics[width=0.5\textwidth]{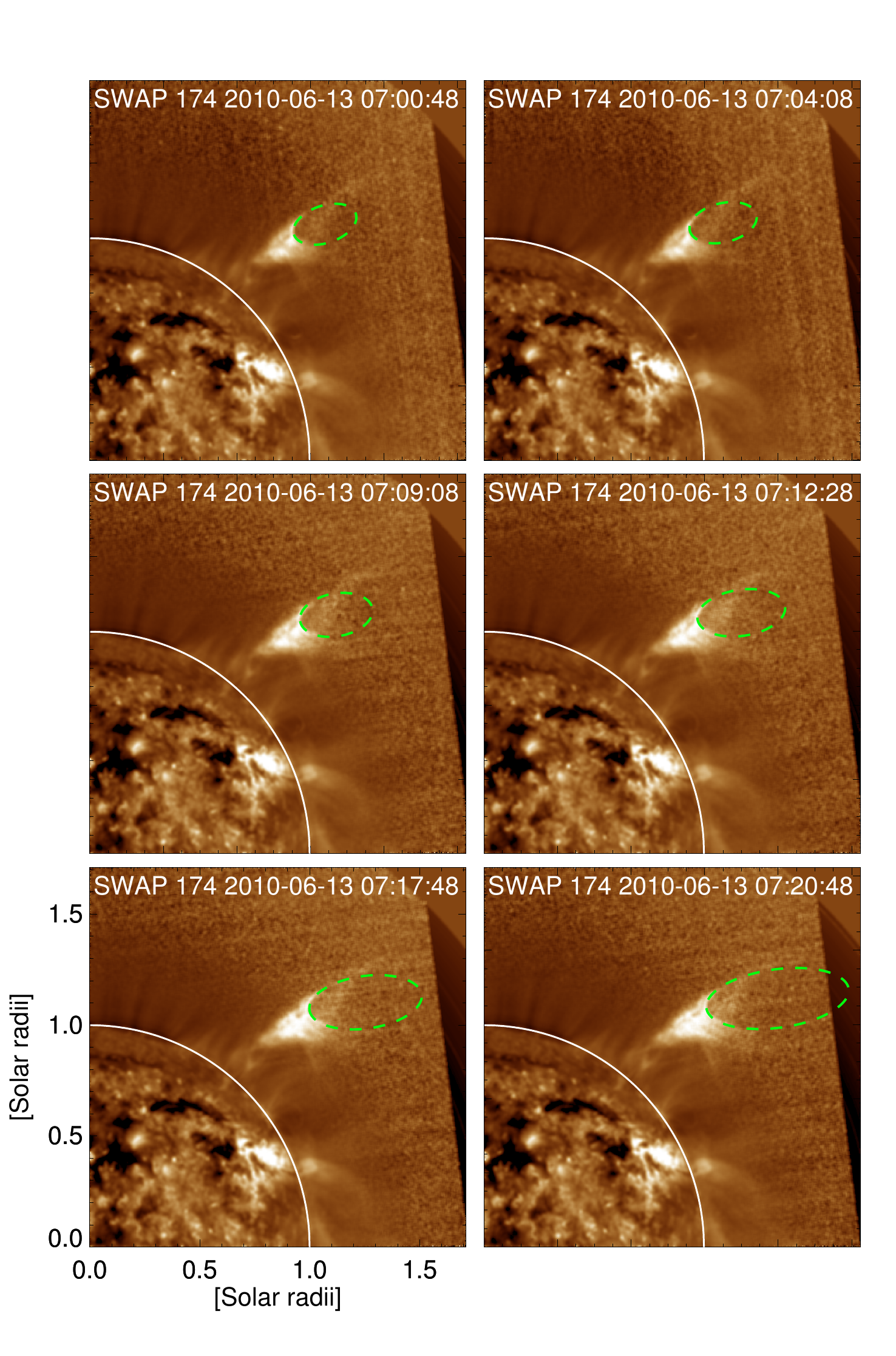}
\caption{Cavity morphology as observed in the SWAP EUV images during the eruptive phase. The solid white line in each panel highlights the position of the solar limb. The green dashed ellipses depict the geometrical fit to the cavity morphology.\label{fig:swap_fit_plot1}}
\end{figure}

The cavity morphology as observed in the SWAP EUV image is best fitted with the elliptical boundary where the cavity centroid reached a height of $\approx$ 0.23 R$_S$ above the solar surface at approximately 00:00 UT on 2010 June 13, prior to the eruption (see Figure \ref{fig:cavity_ellipse_fitting}). 

\begin{figure}[!h]
\centering
\includegraphics[width=0.44\textwidth]{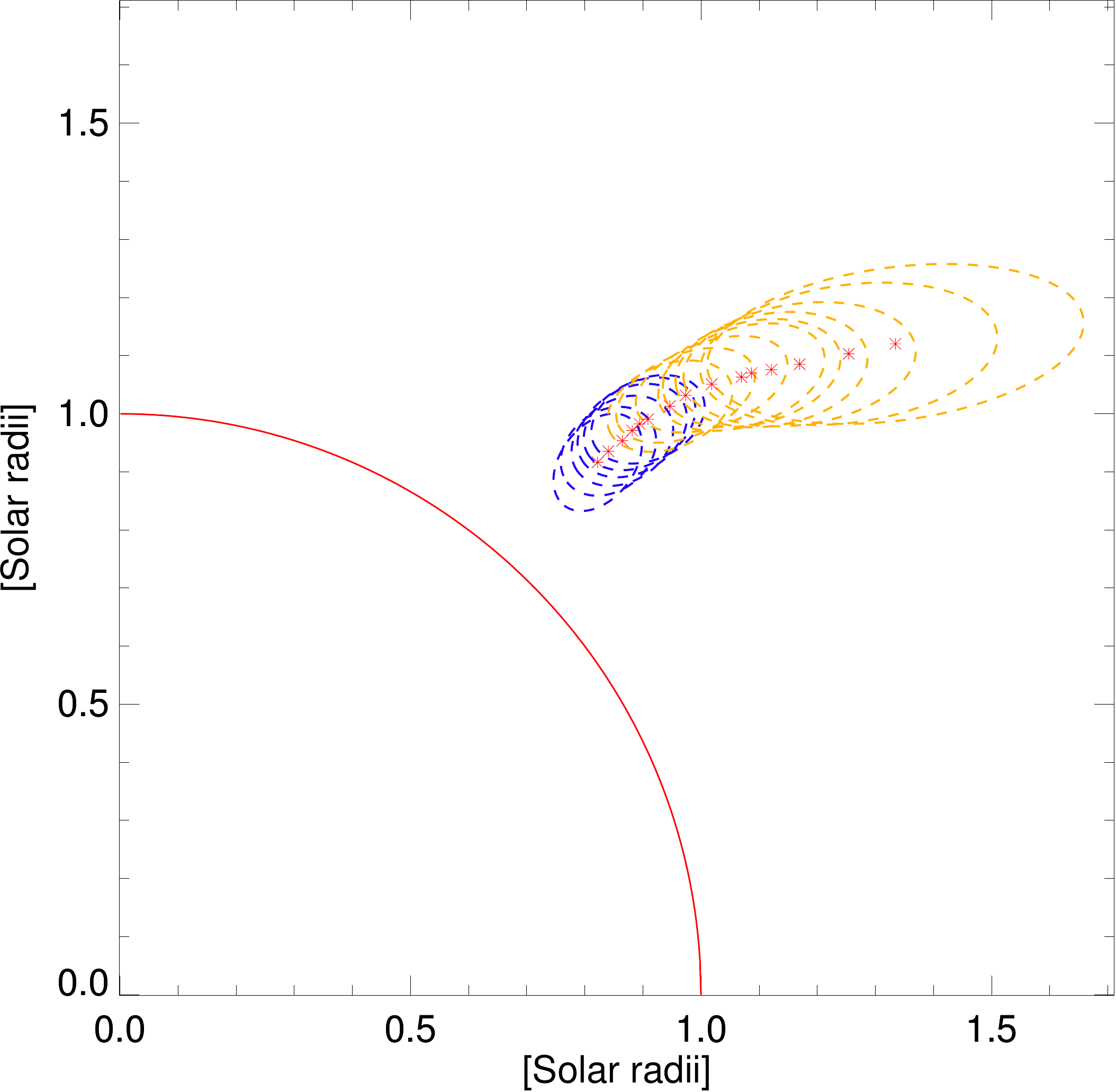}
\caption{Morphological evolution of the cavity system within SWAP FOV during the eruptive phase. The red boundary highlights the solar limb. The blue dotted ellipses denote the geometrical fit to the cavity morphology during the radial motion within 1.3 R$_S$, whereas the yellow dotted ellipses denote the same during the non-radial motion after the deflection at $ \approx $ 1.3 R$_S$. The red asterisks depict the trajectory of the cavity centroid.\label{fig:SWAP_FOV}}
\end{figure}

Although the cavity boundary is clearly detectable in the SWAP composite images, it is difficult to detect the full outer boundary of the cavity in each individual SWAP image. Only the lower boundary of the cavity can be identified due to the emission from the lower lying prominence material which acts as the outer boundary of the cavity. Therefore, we have traced the coordinates along the lower-half boundary of the cavity structure at different times throughout its eruption. These traced coordinates are then fitted with an elliptical geometry, assuming that the upper part of the cavity morphology is symmetrical as the lower-half (see Figure \ref{fig:swap_fit_plot1}). Figure \ref{fig:SWAP_FOV} depicts the morphological evolution of the cavity within SWAP FOV where each of the ellipses represents the geometrically fitted structure of the cavity morphology at different time-steps during its eruptive phase as shown in figure \ref{fig:swap_fit_plot1}.

\begin{figure*}[!h]
\includegraphics[width=\textwidth]{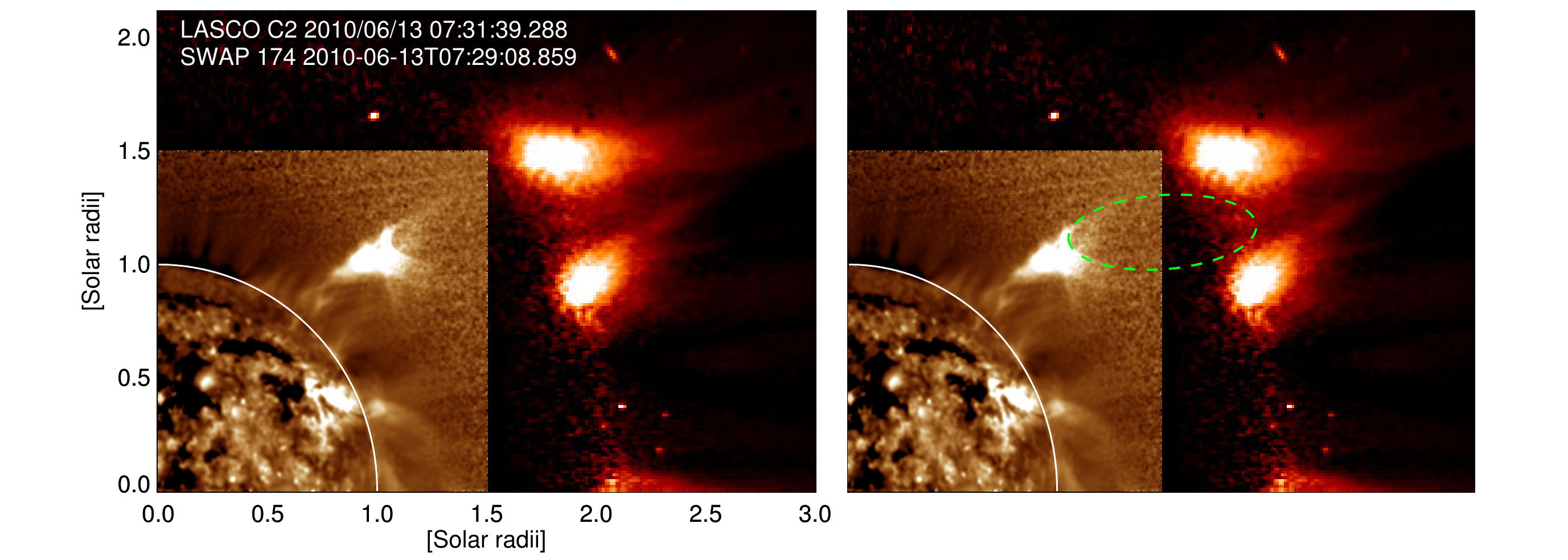}
\includegraphics[width=\textwidth]{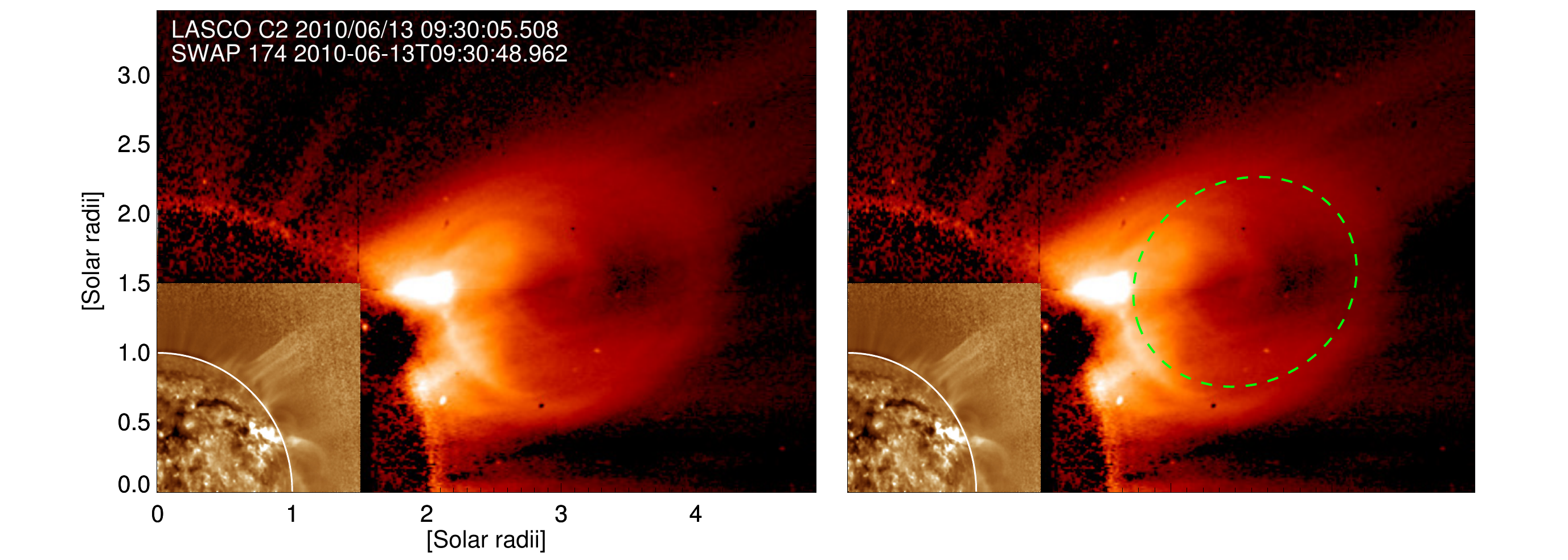}
\includegraphics[width=\textwidth]{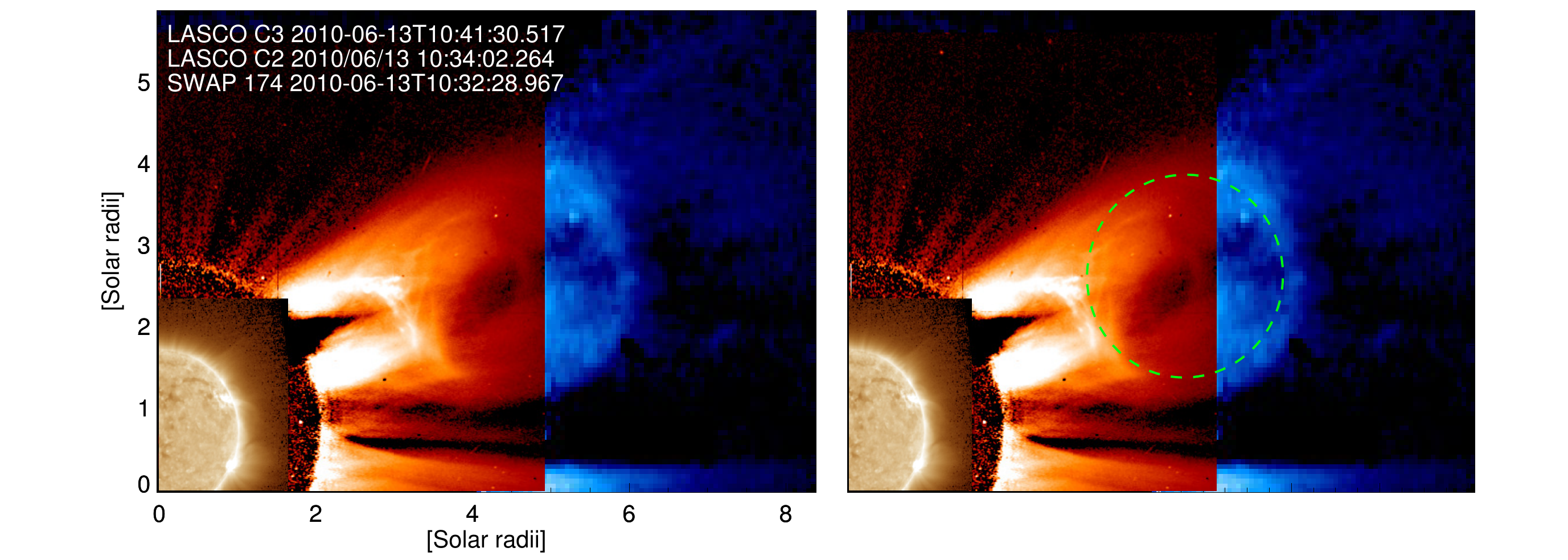}

\caption{Cavity morphology as observed in white-light coronagraphic images of LASCO C2 and C3 observations (red and blue respectively), with SWAP observations superimposed in the center. The white solid lines highlight the solar limb. The green dashed lines show the geometrical fit to the cavity morphology. \label{fig:cavity_euv_wl}}
\end{figure*}

\begin{figure*}[!t]
\centering
\includegraphics[width=0.49\textwidth]{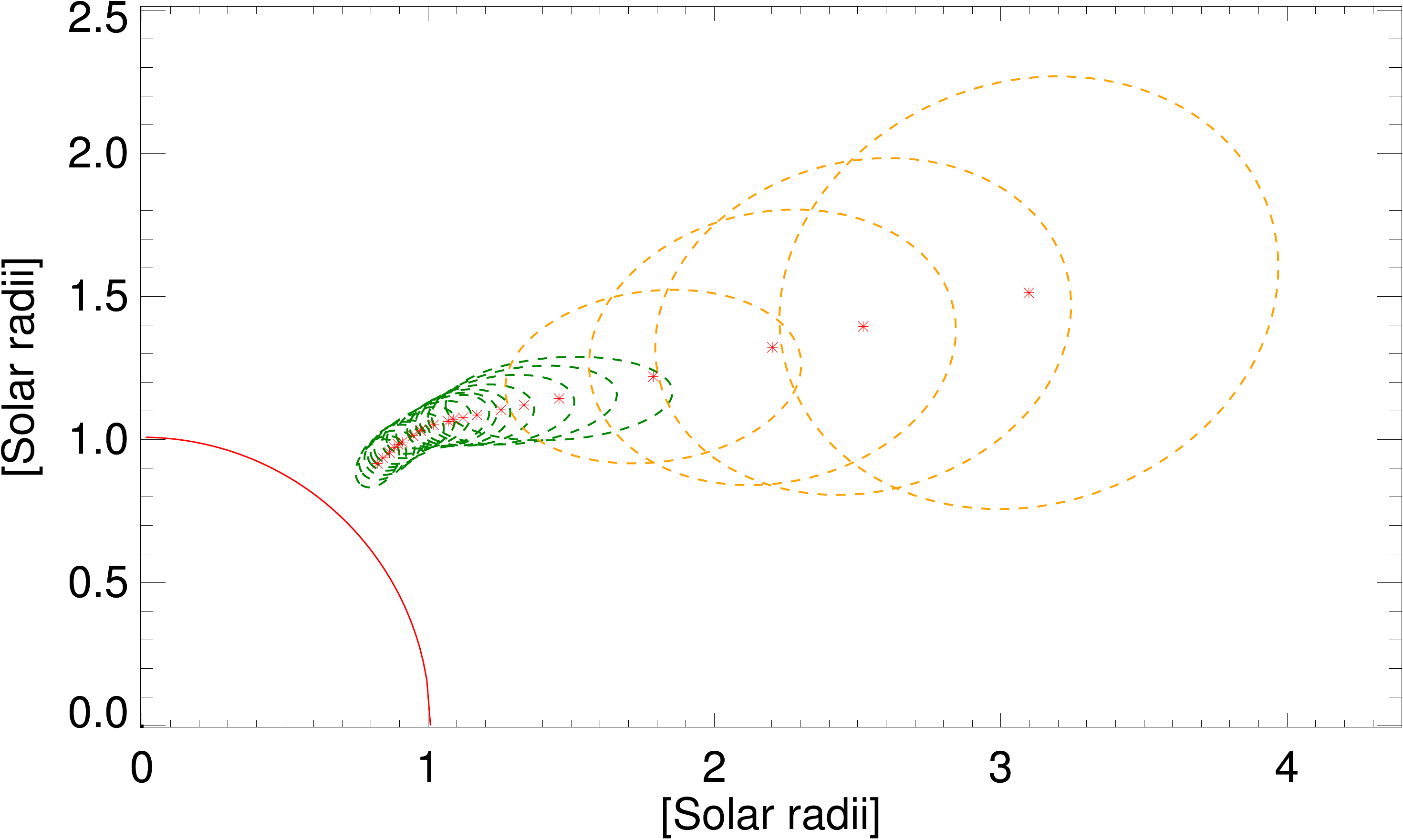}
\includegraphics[width=0.415\textwidth]{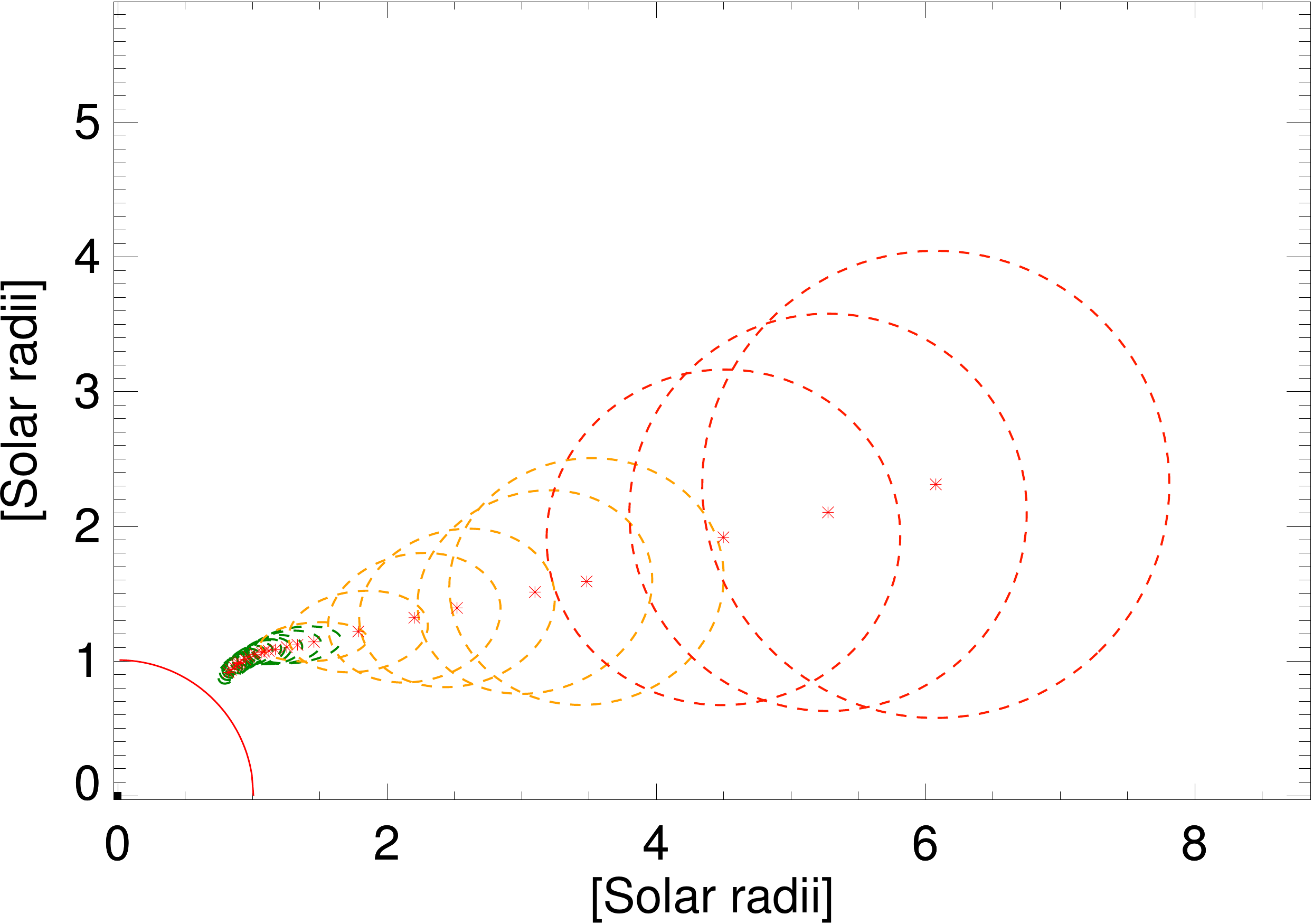}
\caption{The left panel shows the geometrical fit to the cavity morphology observed in the combined FOV of SWAP (green ellipses) and LASCO C2 (yellow ellipses). The right panel shows the geometrical fit to the cavity morphology observed in the combined FOV of SWAP (green dotted ellipses), LASCO C2 (yellow dotted ellipses) and LASCO C3 (red dotted ellipses). The red solid boundary highlights the solar limb. The red asterisks depict the trajectory of the cavity centroid.\label{fig:cavity_track_euv_wl}}
\end{figure*}

\begin{figure*}[!t]
\centering
\includegraphics[width=.92\textwidth]{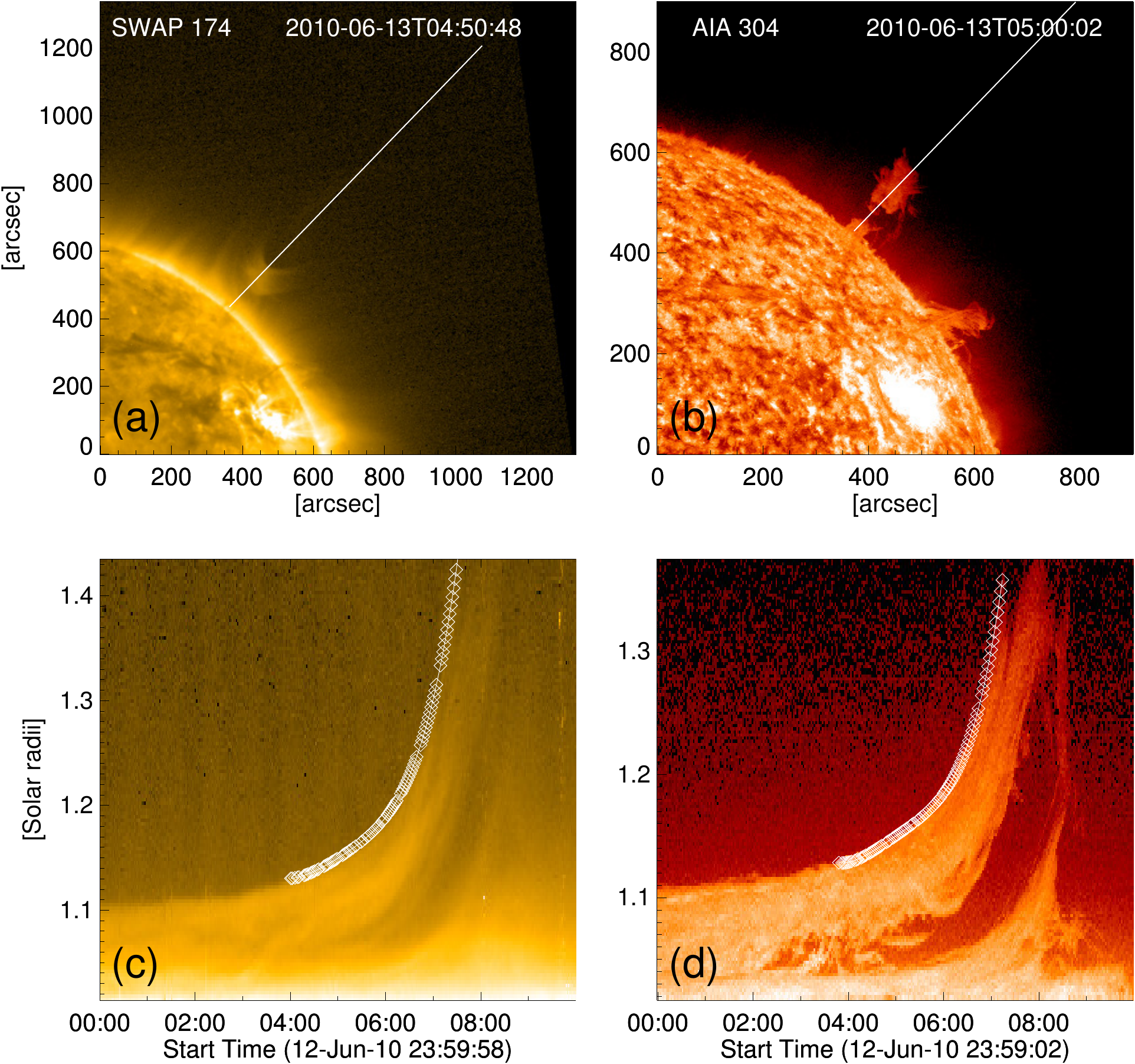}
\caption{The position of the slits superimposed on the SWAP and AIA images (panels (a) and (b) respectively) along with the height-time profiles for the lower-most boundary of the cavity (c) and the top-most part of the prominence (d).\label{fig:filament_cavity}}
\end{figure*}

In order to understand how the EUV cavity at lower corona evolved into the white light cavity associated with the CME, we have combined the SWAP observations with the white light coronagraphic images as captured by the LASCO C2/C3. Figure \ref{fig:cavity_euv_wl} depicts the transformation of the cavity from EUV to white light observations, and its evolution as a three part structure of the associated CME. In order to capture the complete evolution of the cavity, we also apply the geometrical fitting to the white-light cavity morphology. The outer boundary of the dark cavity on the top of the bright filament as observed in LASCO C2/C3 has been fitted up to a height of 8 R$_S$. Beyond 8 R$_S$ the cavity morphology became too diffused to be fitted with certainty. It was possible to track the centroid of the EUV cavity out to 1.8 R$_S$ using the morphological fit to its lower boundary, even when the cavity centroid height exceeded the outer extent of SWAP FOV. Similarly, we have fitted the upper-half of the white-light cavity in LASCO FOV for several frames where the lower part of the cavity was obscured by the occulter and the cavity centroid height resided below 2 R$_S$. Therefore, the small observational gap (1.8 to 2 R$_S$) between the SWAP and LASCO FOV did not affect the continuous tracking of the erupting cavity. The morphological evolution of the coronal cavity starting from its initial centroid height at 1.23 R$_S$ up to 8 R$_S$ can be seen in figure \ref{fig:cavity_track_euv_wl}.

\subsection{Kinematic evolution of the cavity-prominence system}

As discussed in Section \ref{sec:first}, during the quiescent phase the spatial association between the prominence structure and the cavity morphology shows that the prominence material lies exactly at the bottom of the cavity (see Figures \ref{fig:image_east_limb}, \ref{fig:image_stereo_limb} and \ref{fig:swap_cavity}). In order to investigate whether this spatial relationship is also maintained during the eruptive phase, we have separately tracked the top boundary of the prominence material as observed in AIA 304 \AA\ channel and the bottom boundary of the coronal cavity as observed in SWAP EUV images. This was done by placing a slit at a position angle of 317$^\circ$ on both the SWAP and AIA images, and stacking the evolution of the cavity and filament boundary with time within the respective slits. Figure \ref{fig:filament_cavity} shows the position of the slits (top panels) in the SWAP EUV and AIA 304 \AA\ images respectively. The bottom panels of figure \ref{fig:filament_cavity} illustrate the evolution of the bottom boundary of the cavity (panel (c)) and the top boundary of the filament (panel (d)). In order to compare these two height-time profiles, we have selected the points with proper coordinate informations along the two height-time curves as marked by the square diamond shaped symbols drawn on the two time-slice diagrams (Figure \ref{fig:filament_cavity}). These selected points along the two height-time curves are then over-plotted in figure \ref{fig:combine_plot}. The error bars ($\pm$ 0.006 R$_S$) are a measure of the uncertainty in selecting the points along the curve boundary.

\section{Results} \label{sec:results}
\subsection{Spatial relation between the EUV cavity and the associated prominence}
\begin{figure}[!h]
\centering
\includegraphics[width=0.47\textwidth]{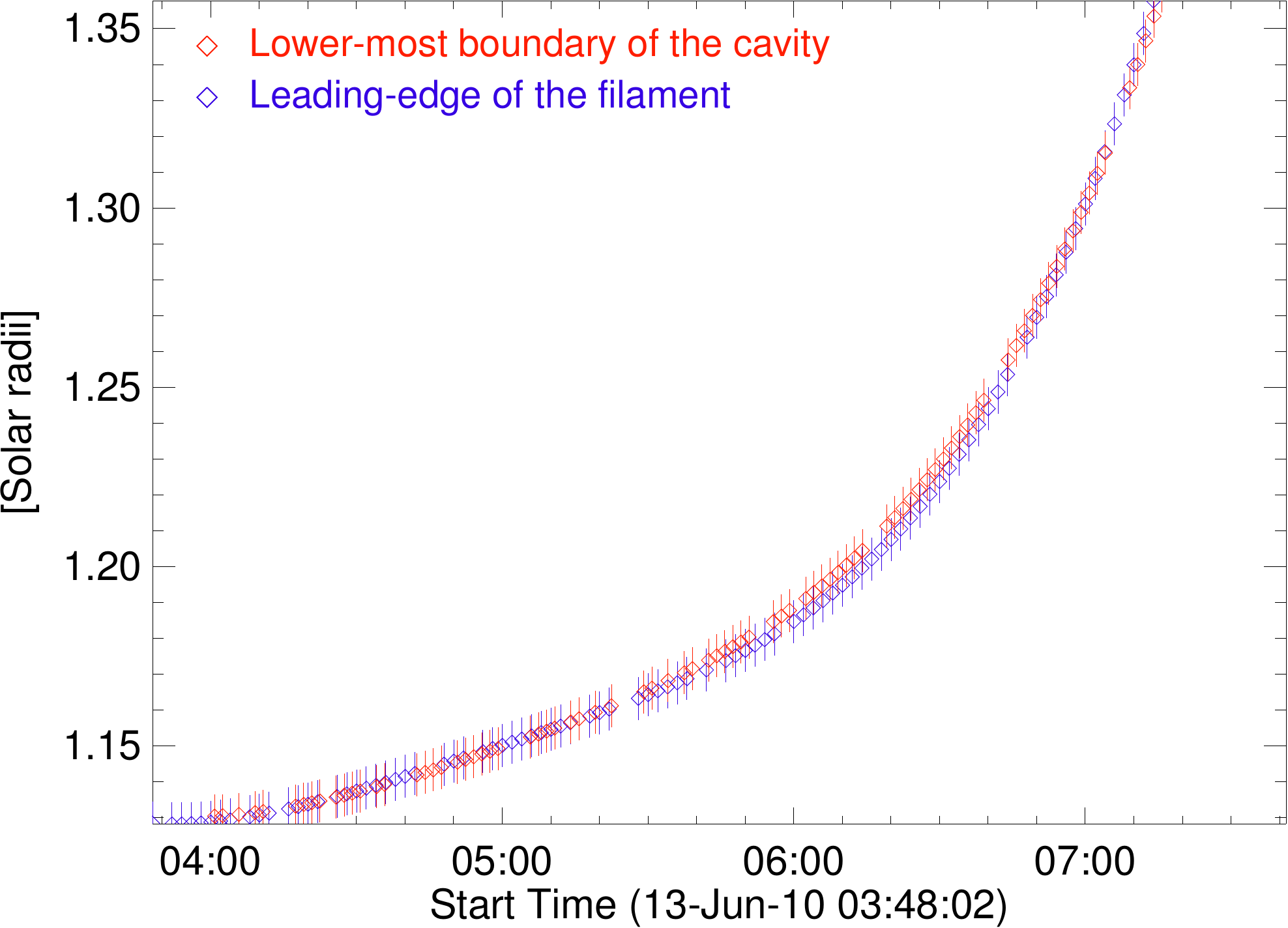}
\caption{The height-time profiles of the lowermost boundary of the cavity (red) and the leading edge of the filament (blue), as recorded in Figure \ref{fig:filament_cavity}.\label{fig:combine_plot}}
\end{figure}

Combining the observations of the erupting cavity and the associated prominence structure we have studied the spatial relationship between them in lower corona during both the quiescent and eruptive phases. Figure \ref{fig:combine_plot} clearly shows that the height-time profiles of the filament top and the cavity bottom boundary coincide with each other. This indicates that throughout the eruptive phase the cavity is located exactly on the top of the prominence which is in agreement with the understanding of the classical three part structure of CMEs.      
  
\begin{figure*}[!ht]
\centering
\includegraphics[width=\textwidth]{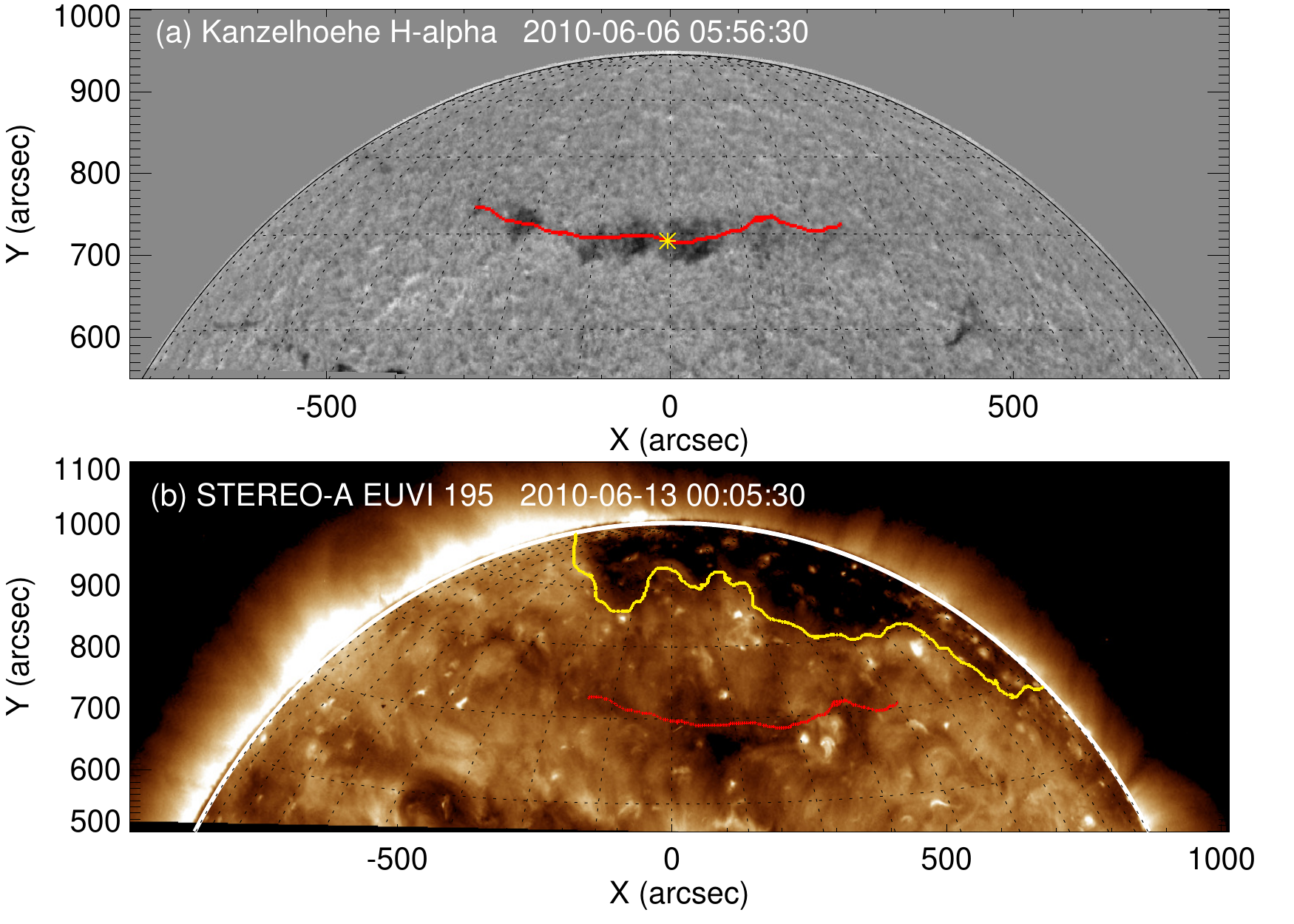}
\caption{H$\alpha$ image of the polar crown filament on 2010 June 6 (a). The red solid line, drawn over the image, depicts the approximate position of the filament channel on the solar disk. The yellow star mark approximately locates the central part of the long filament channel. STEREO-A EUVI 195 \AA\ image on 2010 June 13 (b). The yellow solid line denotes the boundary of the northern polar coronal hole. The red solid line in panel (b) indicates the location of the same filament channel as shown in panel (a).
\label{fig:gong_h_alpha}}
\end{figure*}

\subsection{Cavity morphology during the quiescent phase}
  
By using the multi-spacecraft observations from SDO, STEREO and PROBA2 we have tracked the gradual evolution of the stable EUV cavity at different times during its passage across the solar disk. 

However, in order to relate the different evolutionary phases of the quiescent cavity it is important to track the same portion of the large 3D structure associated with it. Figure \ref{fig:gong_h_alpha} shows the location of the associated H-alpha filament channel on 6th June 2010, when it was close to the solar disc center. The longitudinal extent of the filament channel was approximately 60$^\circ$ and the Carrington heliographic longitude of the central part (indicated by the yellow star mark in panel (a) of Figure \ref{fig:gong_h_alpha}) of this long filament channel was 132 $\pm$ 4$^\circ$. Noticeably, the Carrington longitude of the quiescent cavity system on 2010 May 30, June 4, June 7 and June 13 as observed sequentially by the SDO, STEREO-A, STEREO-B and PROBA2 was 136$^\circ$, 132$^\circ$, 134$^\circ$ and 132$^\circ$ respectively (see Table \ref{tab:table1}) which was nearly the same as the Carrington longitude of the central part of the filament channel. Moreover, as the orientation of the filament channel was nearly horizontal with respect to the solar equator, the associated flux-rope possessed a close to azimuthally symmetric structure with respect to the solar rotational axis. Therefore, all the observations depicting the different phases of the quiescent cavity capture the same part of the large 3D structure and hence reveal the true evolution of the cavity during its long-lived quiescent phase.     

The geometrical fit to the cavity structure reveals that the cavity boundary had a near circular morphology when it appeared on the east solar limb, and still had this shape when it was positioned near disk centre, as seen from the POS EUVI observations of STEREO A and B (see Figures \ref{fig:image_east_limb} and \ref{fig:image_stereo_limb}). However, the cavity morphology evolved into a more elliptical shape as it approached the west solar limb, prior to  eruption (see Figure \ref{fig:swap_cavity}). Apart from the morphological change, the quiescent cavity also undergoes a slow rise and expansion phase. As the cavity rotated across the solar disk from east to west solar limb, its diameter increased from 0.016 $\pm$ 0.002 to 0.09 $\pm$ 0.02 R$_S$ and the cavity centroid height raised from 1.10 to 1.23 $\pm$ 0.002 R$_S$ (see Table \ref{tab:table1}) before it finally erupted from the west solar limb.

\subsection{Cavity dynamics during the eruptive phase}
Using the large FOV of SWAP we have sequentially captured the evolution of the erupting cavity morphology in the lower corona up to 1.7 R$_S$. Figure \ref{fig:SWAP_FOV} presents a clear depiction of a significant non-radial motion exhibited by the cavity at about 1.3 R$_S$ where its position angle changed from approximately 310$^\circ$ to 270$^\circ$. In general, the equator-ward deflection of CMEs is believed to be due to the influence of polar coronal holes \citep{cremades,panasenco2011}. However, the EUV images obtained from the AIA and SWAP on 2010 June 13 do not show any signature of a polar coronal hole in the northern hemisphere which may be due to the line-of-sight tilt of the solar magnetic axis. Therefore, in order to identify the presence of a polar coronal hole we have used the EUV observations of Sun on the same day from the perspective of STEREO-A (see the position of STEREO-A in Figure \ref{fig:st_combine_new}).

Panel (b) in Figure \ref{fig:gong_h_alpha} clearly shows the presence of a northern polar coronal hole on 2010 June 13 as observed by STEREO-A EUVI in 195 \AA\ passband. In order to identify the location of the filament channel on the EUVI 195 \AA\ image, we first computed the Carrington longitude and latitude of the points along the H$\alpha$ filament channel (indicated by the red solid line in panel (a) of Figure \ref{fig:gong_h_alpha}) as observed on 2010 June 6. Using the information of the Carrington heliographic coordinates of these points, the same filament channel has been overlaid on the STEREO-A EUVI 195 \AA\ image. The location of the filament channel in the vicinity of the northern polar coronal hole reveals that the open magnetic field-lines originating from the polar coronal hole might have channeled the trajectory of the erupting cavity towards the heliospheric current-sheet. Therefore, the cavity deflected from a higher latitude to a lower one.

By combining the observations from SWAP and LASCO C2/C3 we have studied the association between the EUV and white light observations of the cavity. Figure \ref{fig:cavity_track_euv_wl} shows that during the eruptive phase the cavity morphology gradually evolved from an initial elliptical shape to a semi-circular one (see left panel of Figure \ref{fig:cavity_track_euv_wl}) within the FOV of LASCO C2. As the cavity evolved further, it became almost circular after about 4 R$_S$. Noticeably, after the initial deflection at about 1.3 R$_S$ the cavity maintained the same position angle (270$^\circ$) throughout the rest of its propagation trajectory.     
\subsection{Nature of expansion of the erupting cavity}
In order to investigate whether the cavity evolution was self-similar or not, we have plotted the evolution of the cavity aspect-ratio with respect to the cavity centroid height. The numerator of the aspect-ratio is the length of the semi-major axis of the cavity ellipse and the denominator is the distance from Sun center to the cavity centroid along the non-radial trajectory of the cavity propagation.   

\begin{figure*}[!ht]
\centering
\includegraphics[width=.7\textwidth]{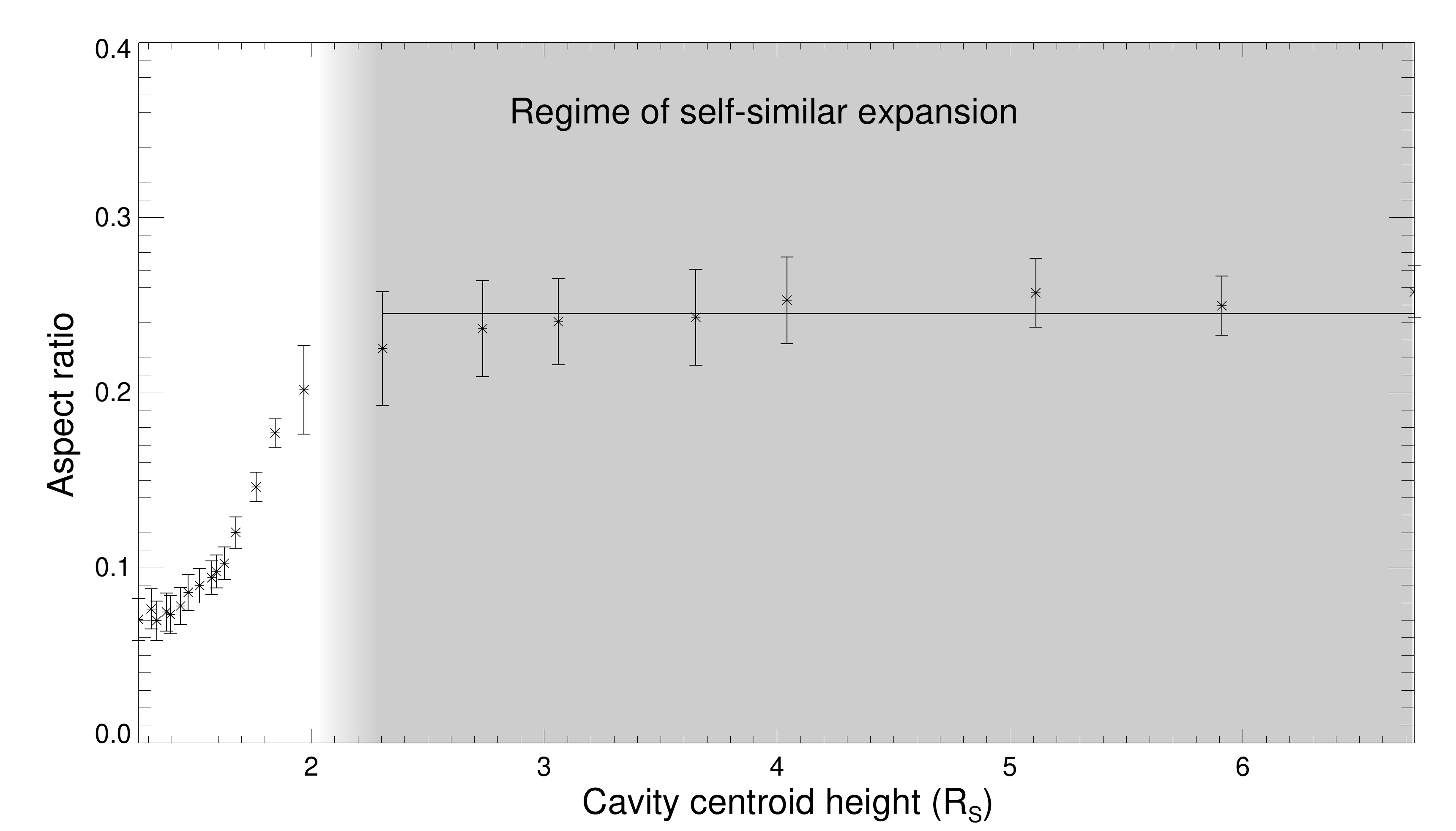}
\hfill
\includegraphics[width=.7\textwidth]{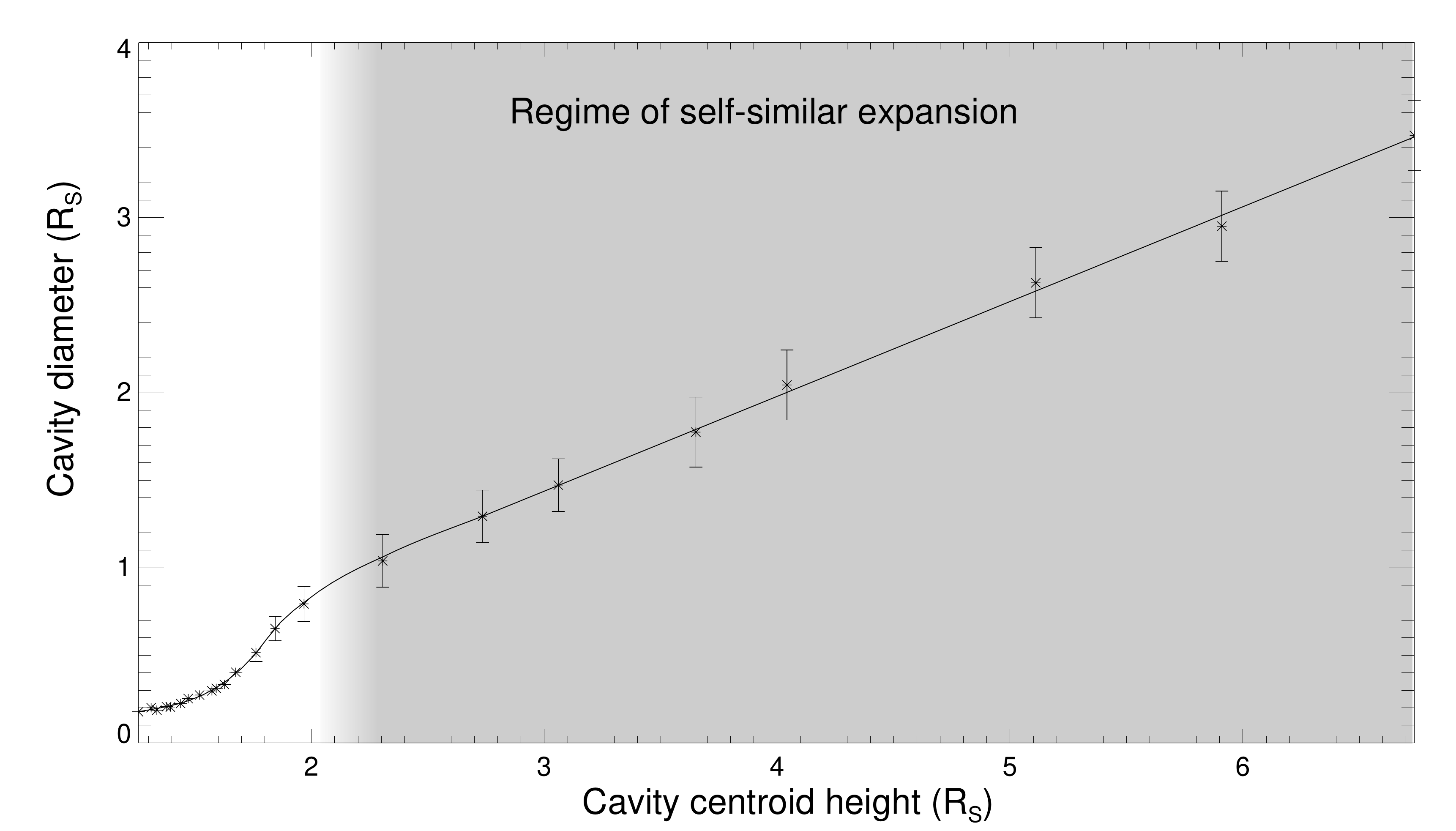}
\hfill
\caption{The evolution of the cavity width to centroid height aspect-ratio (top panel). Evoloution of the length along the semi-major axis of the cavity with respect to the cavity centroid height (bottom panel).\label{fig:self_similar}}
\end{figure*}

\begin{figure*}[!t]
\centering
\includegraphics[width=\textwidth]{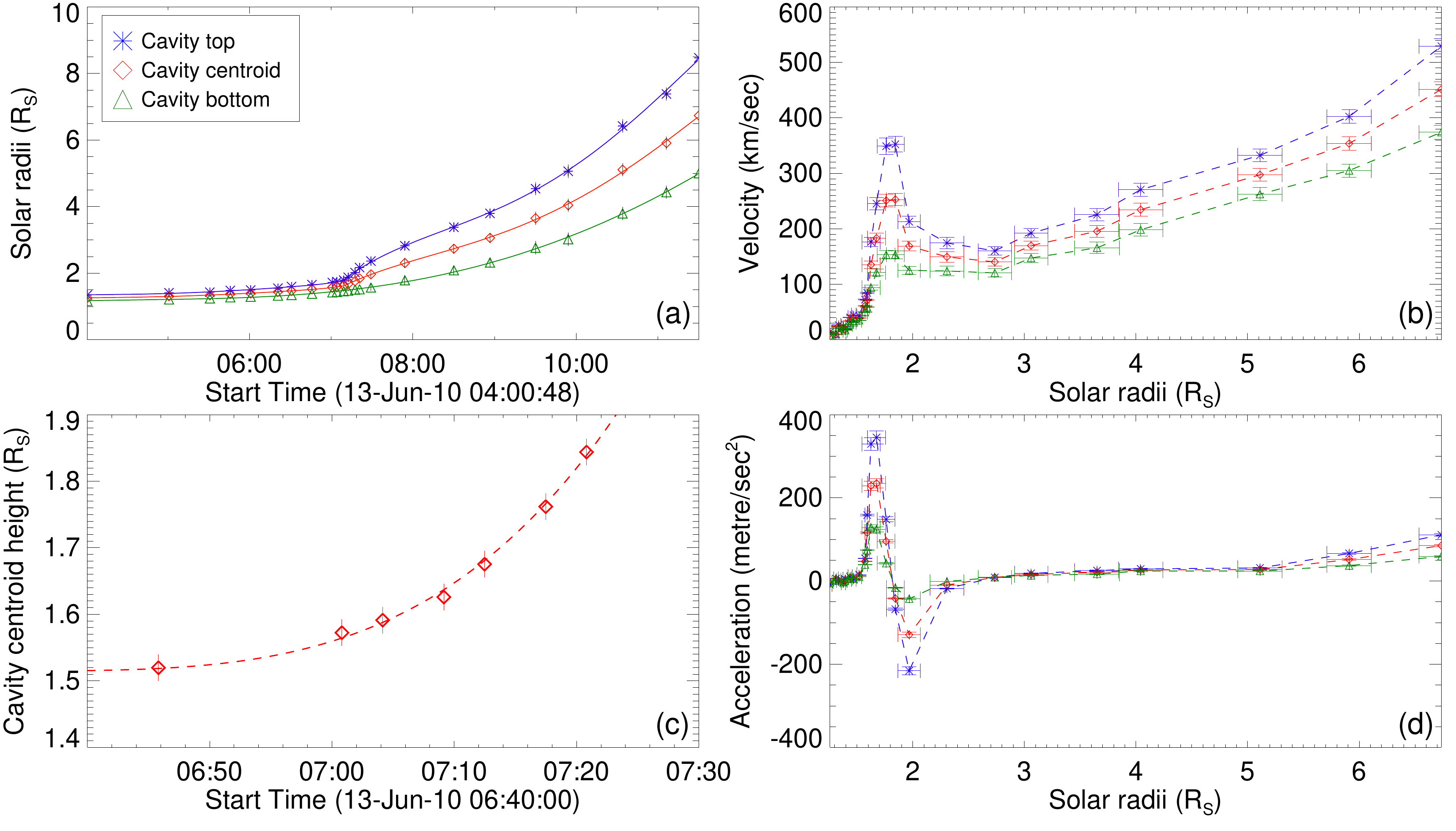}
\caption{Height-time profiles for the cavity top, cavity centroid and cavity bottom (a). Velocity profiles for the cavity top, cavity centroid and cavity bottom (b). Polynomial fit to the cavity centroid height during the time interval 06:40 to 07:30 UT on 2010 June 13 (c). Acceleration profiles for the cavity top, cavity centroid and cavity bottom (d). \label{fig:ht_vel}}
\end{figure*}

The upper panel of figure \ref{fig:self_similar} shows the evolution of the cavity width to centroid height aspect-ratio, which can be seen to gradually increased to 0.25 $\pm$ 0.03 at 2.2 $\pm$ 0.2 R$_S$, after which it became approximately constant for the remainder of its propagation. This clearly shows that within 2.2 $\pm$ 0.2 R$_S$ the cavity exhibited non-self similar expansion whereas beyond 2.2 $\pm$ 0.2 R$_S$ it enters into a regime of self similar expansion. The non-self similarity in the cavity evolution is also reflected in its expansion profile in the lower corona. The bottom panel of figure \ref{fig:self_similar} shows that the cavity diameter (length along the major axis of the ellipses fitted to the cavity morphology) expands linearly after 2.2 $\pm$ 0.2 R$_S$, whereas it evolves in a different fashion within the non-self similar regime in lower corona. Noticeably, this critical height (2.2 $\pm$ 0.2 R$_S$) resembles the radius of source surface (2.5 $\pm$ 0.25) where the coronal magnetic field-lines are believed  to become radial \citep{Hoeksema}.
\subsection{Kinematic evolution of the cavity}
Investigating the kinematics of the erupting cavity we find that the individual height-time profiles of cavity top, centroid and bottom part evolve in a different manner due to the internal expansion of the cavity structure (panel (a) of Figure \ref{fig:ht_vel}). Interestingly, the velocity profiles shown in figure \ref{fig:ht_vel} depict that the erupting cavity undergoes two distinct phases of kinematic evolution. 

During the first phase, the speed of the top, centroid and bottom part of the cavity quickly reached approximately 350, 250, 150 km/sec respectively within $\approx$ 1.8 R$_S$. However, the second phase of the kinematic evolution starts after 2 R$_S$ where the speed of the three different parts of the cavity increases more gradually in comparison to the first kinematic phase. Noticeably, the cavity centroid exhibited the peak acceleration at 1.67 $\pm$ .08 R$_S$ where the top, center and bottom part of the cavity attained the accelerations of 345 $\pm$ 15, 234 $\pm$ 11 and 124 $\pm$ 6 m/sec$^2$ respectively. This impulsive phase of acceleration is believed to be governed by the Lorentz self-force where the outward magnetic pressure dominates over the external and/or internal magnetic tension force \citep{Byrne2010}. Importantly, the peak acceleration height (1.67 $\pm$ 0.08 R$_S$) of the cavity obtained in this study is in agreement with the mean value (1.72 R$_S$) of that found for the filament associated CMEs studied by \citet{Bein2012}. However, after 2 R$_S$ the average acceleration reduces to below 50 m/sec$^2$ which is believed to be the ``residual acceleration phase'' of the CME where the Lorentz self-force undergoes the declining phase and the flux-rope dynamics become strongly dependent on the drag force \citep{chen2003}.      

In order to initiate the first phase of the kinematic evoluton, several triggering mechanisms have been proposed (see the review by \cite{PFChen2011}). Catastrophic loss of equilibrium, ideal MHD instabilities, tether-cutting, magnetic breakout, triggering by flux-emergence or cancellation and mass drainage are believed to be one of the possible driving mechanisms for triggering the initial acceleration phase. Interestingly, by fitting the height-time profiles of the kinematic evolution with a power-law polynomial and by comparing it with the results obtained from different numerical simulations, one may get clues to the underlying eruption mechanism \citep{Torok2005,Williams2005,Marilena2013}. 

\citet{Schrijver2008} have shown that the rapid-acceleration phases of erupting prominences are best characterized by a height dependent function, $h(t) \propto t^m$, where h(t) is the instantaneous height at any time t and m is the power law exponent.
 Using this power law function, we have fitted the height-time profile of the cavity centroid during its impulsive acceleration phase in between 06:40 to 07:30 UT on 2010 June 13. The red dashed line in panel (c) of figure 15 shows the curve of best fit obtained by using $m=3.6$. Noticeably, an exponent value (m) close to 3 corresponds to the torus-instability scenario \citep{Schrijver2008}. Therefore, our results suggest that the impulsive acceleration phase was most likely driven by the torus instability.

\subsection{Eruption mechanism of the coronal cavity in the context of torus instability}
\setlength{\tabcolsep}{3pt}
\begin{deluxetable}{ccc}[!hb]

\tablecaption{Temporal evolution of decay index \label{tab:table2}}
\tablehead{
\colhead{Observation time} & \colhead{Decay index at the} &\colhead{Decay index at the} \\
\colhead{yyyy/mm/dd hh:mm:ss} & \colhead{top of the filament } &\colhead{cavity centroid} 
}

\startdata
2010/05/30 02:32:02 & 0.75 $\pm$ .07& 0.80 $\pm$ .06\\
2010/06/04 21:25:30 & 0.7 $\pm$ .2& 0.7 $\pm$ .2\\
2010/06/07 14:26:02 & 0.7 $\pm$ .1& 0.8 $\pm$ .1\\
2010/06/13 04:00:00 & 1.2 $\pm$ .1& 1.3 $\pm$ .1\\
\enddata

\end{deluxetable}

In order to understand the pre eruptive stability conditions for quiescent cavities and the triggering mechanisms for those structures to erupt, we have examined the role of the background magnetic field in the context of torus instability. In order to obtain the information of the overlying magnetic field we have used the Potential Field Source Surface (PFSS) extrapolation code \citep{derosa} available in SolarSoft packages. As the minimum time-cadence of the available extrapolated fields from PFSS model is 6 hours, we have used the extrapolations carried out within the minimum temporal window from the observing time of the cavity. For each of the four different days when the cavity was observed sequentially on the east-limb, POS of STEREO A, STEREO B and west-limb, the decay index of the overlying magnetic field has been evaluated along the radial direction through the cavity centroid lying on the respective POS. In order to calculate the decay-index we have used the following formulation

$$\rm{decay\ index\ [n]}= - \frac{\partial {\rm log}(B_{\rm h})}{\partial {\rm log}(h)} \ ,$$ where h is the height above the solar photosphere and $B_{\rm h}$ is the horizontal component of the external magnetic field obtained from the PFSS extrapolations \citep{Torok2005}. 

The critical value (n$_{critical}$) of decay index for the onset of the torus instability is believed to be close to 1.5 
\citep{Kliem2006}. However, depending on the typical range of current-channel thickness expected in the corona, n$_{critical}$ can vary within a range between 1.2 to 1.5 \citep{Demoulin2010}. From the observational studies on prominence eruptions, the value of n$_{critical}$ (0.9 to 1.1) at the top of the prominence has been found to be less than that found in the MHD simulations \citep{ F2013,Zuccarello_2014,McCauly}. Addressing this discrepancy between the models and observations, \citet{Zuccarello2016} performed a set of MHD simulations and have shown that the value of n$_{critical}$ 
computed at the flux rope's axis is 1.4 $\pm$ 0.1, while at the height of the top of the prominence this value is 1.1 $\pm$ 0.1 . These results suggest that the choice of location (filament top or the cavity centroid) to evaluate the decay index is important for comparing the observational and theoretical values of n$_{critical}$. The temporal evolution of the altitude of filament top and the cavity centroid, as presented in the current paper, allows us to address the above mentioned issue observationally. 

\begin{figure*}[!ht]
\centering
\includegraphics[width=.48\textwidth]{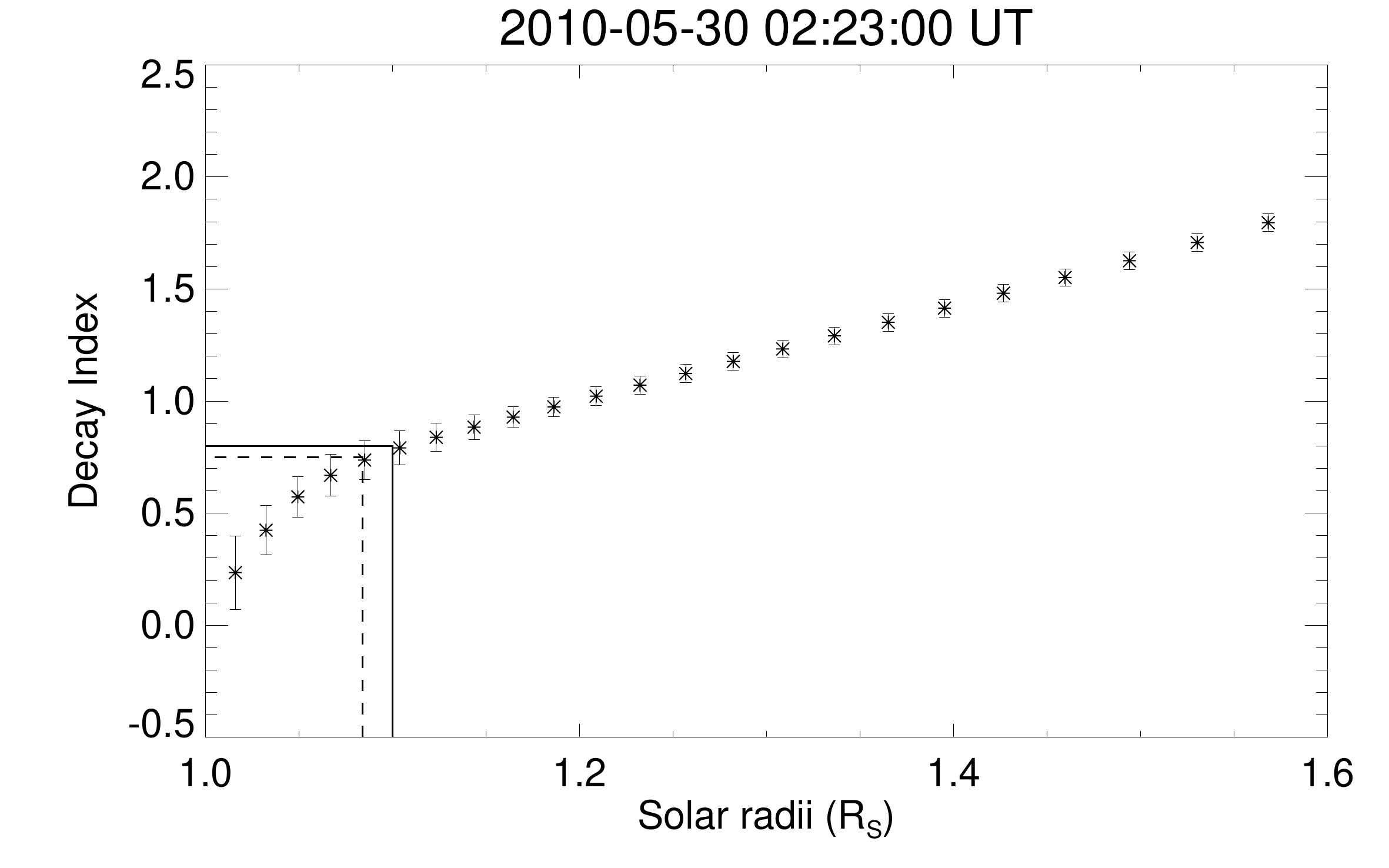}
\includegraphics[width=.48\textwidth]{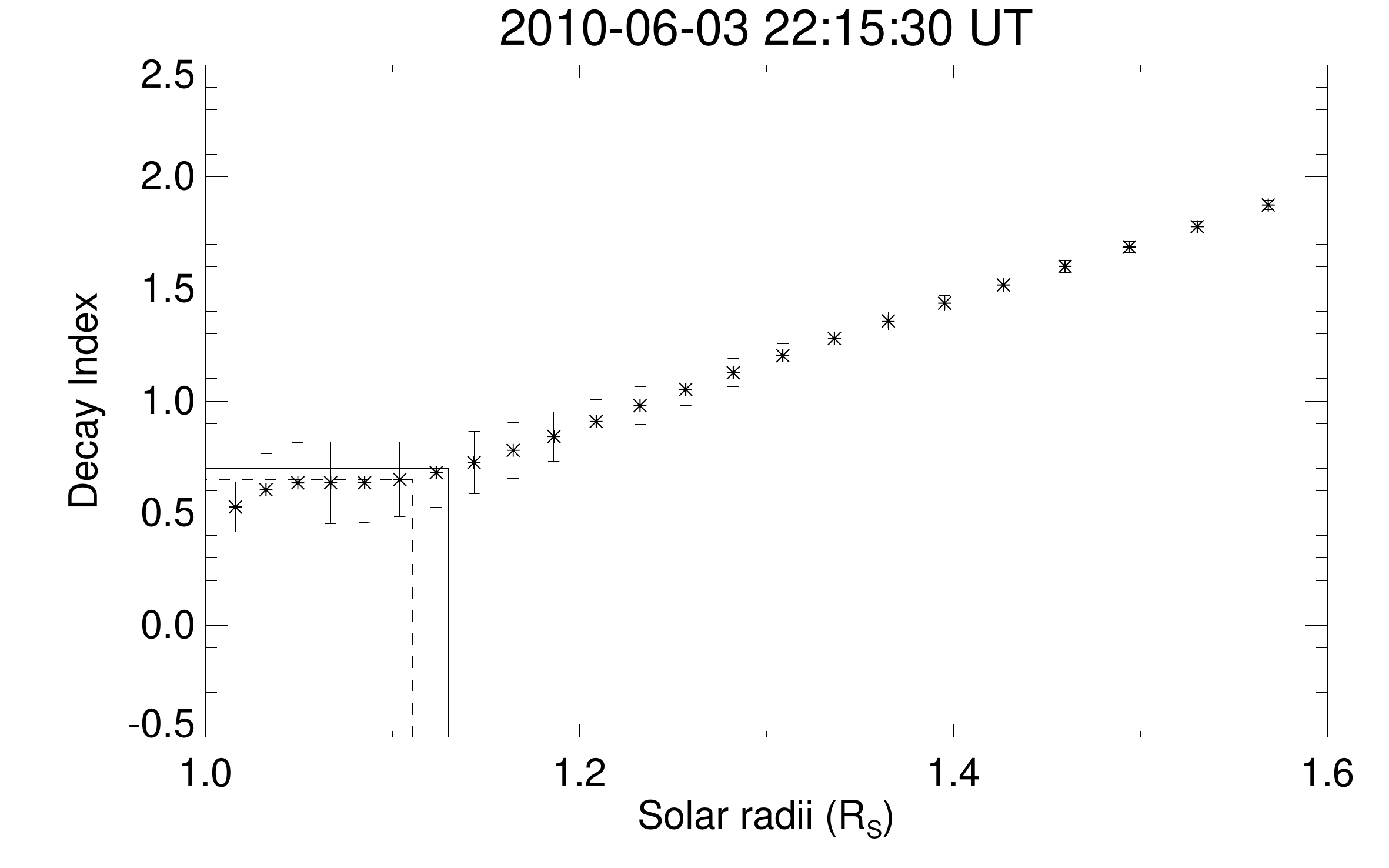}
\hfill
\includegraphics[width=.48\textwidth]{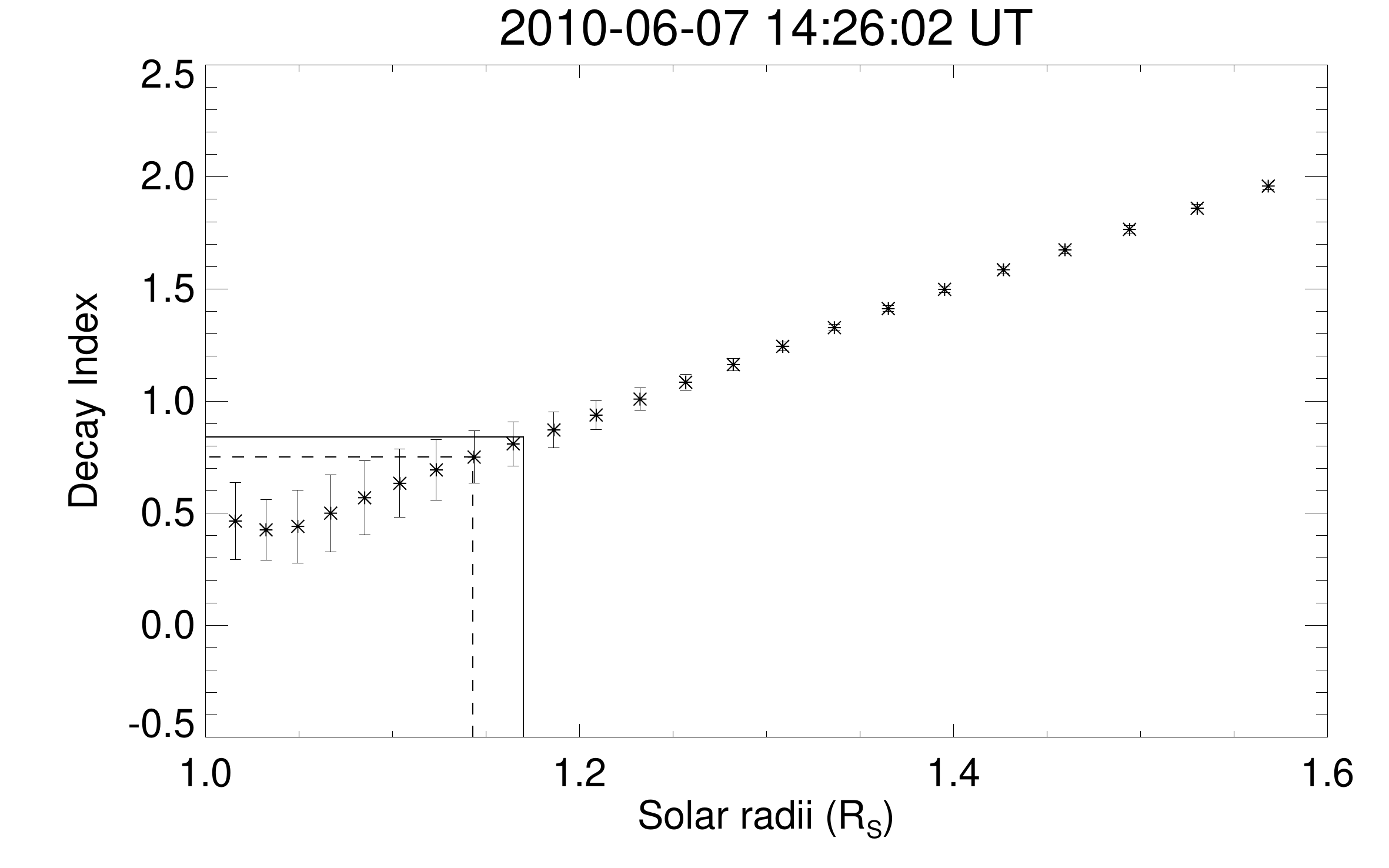}
\includegraphics[width=.48\textwidth]{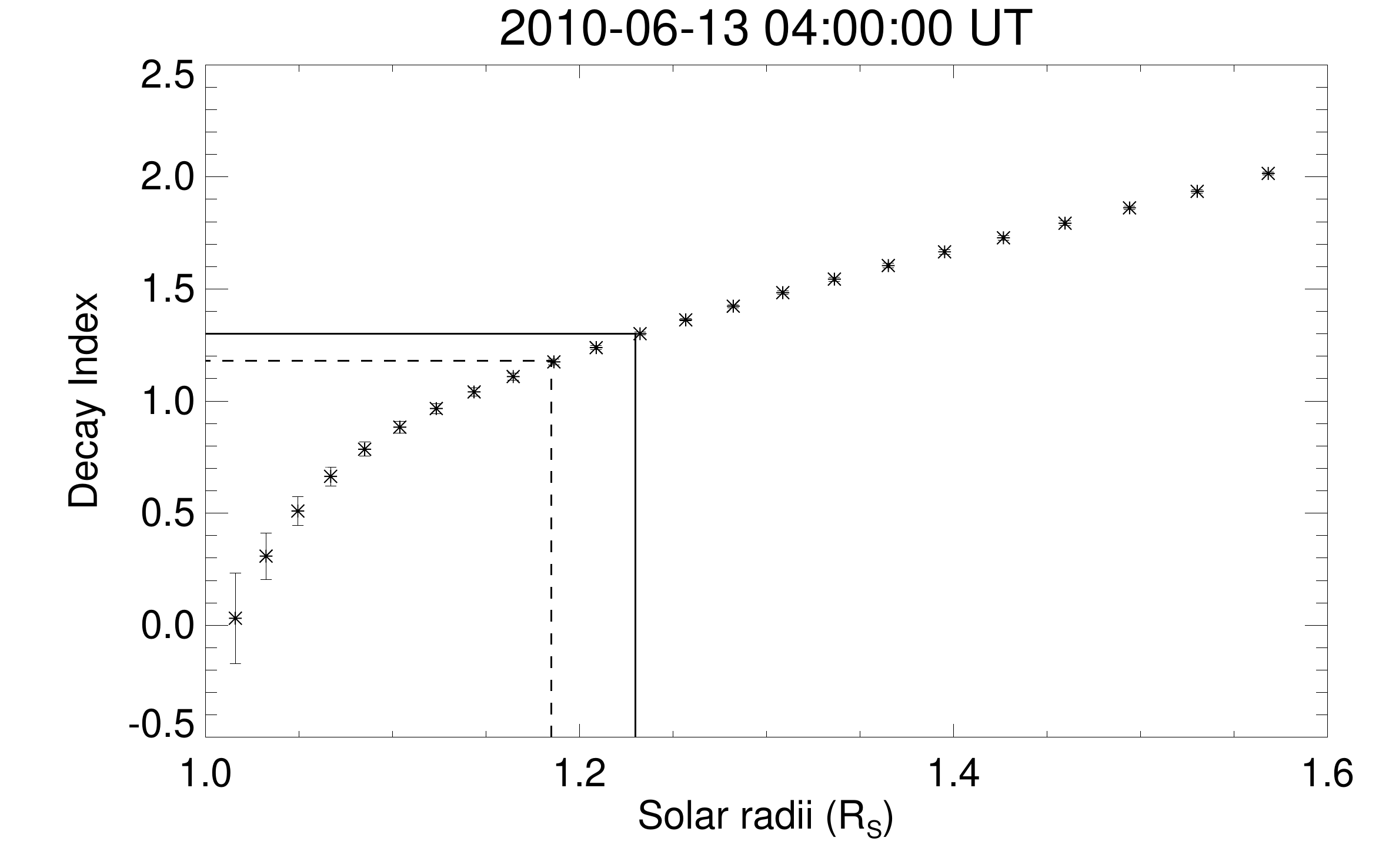}
\caption{Decay index profiles for the different phases of the quiescent cavity. The black solid and dashed lines mark the decay index values at the cavity centroid and the top of the filament respectively. \label{fig:decay_index}}
\end{figure*}

Figure \ref{fig:decay_index} shows the four decay index profiles as calculated for  the four different POS observations of the cavity along 90$^\circ$ E, 18$^\circ$ E, 20$^\circ$ W and 90$^\circ$ W in Stonyhurst heliographic coordinates  \citep{Thompson2006} on 2010 May 30, June 4, 7 and 13 respectively. The overplotted black dashed and solid lines mark the decay index values at the top of the filament and the cavity centroid respectively, which are listed in table \ref{tab:table2}.

The decay index values for the first three positions of the cavity during the observational period from 2010 May 4 to 2010 June 7  indicate that both the filament top and the cavity centroid reside well below the critical decay index limit, while the cavity was in a stable condition (see Table \ref{tab:table2}). However, the decay index value (1.3 $\pm$ .1) at the cavity centroid height reached the critical limit 1.4 $\pm$ .1 for the onset of torus instability prior to the eruption on 2010 June 13. Noticeably, the altitude of the filament top also reached the decay index value (1.2 $\pm$ .1) which resembles the ``apparent" critical value (1.1 $\pm$ .1) of decay index at the filament top as obtained from the simulational results of \citet{Zuccarello2016}. Importantly, the good agreement between the observational (1.3 $\pm$ .1) and simulational (1.2 to 1.5) results for the critical decay index value at the cavity centroid height suggests that the cavity centroid should be used as the preferable location to evaluate the decay index value in order to reduce the discrepancy between the observational and theoretical critical limit for n$_{critical}$. 

Furthermore, our results also suggest that the decay index value at the cavity centroid height can be used as a good indicator for determining whether the cavity will result in an eruption or not.



\section{Discussion and Conclusions} \label{last_sec}

For a two week long quiescent phase of the coronal cavity, we have studied its morphological evolution by tracking the same portion of a large 3D structure associated with it. Combining the observations from the SWAP, LASCO C2 and C3, we have also captured its evolution during the eruptive phase which enables us to associate the erupting EUV cavity with its white-light counterpart as observed in the outer corona. The evolution of the cavity morphology at different stages of its quiescent and eruptive phases reveals the underlying eruption mechanism and the role of magnetic forces governing the cavity dynamics.     

Observations from the different vantage points during the quiescent phase reveal that the cavity morphology maintained a close to circular shape as it rotated across the solar disk. Eventually, the circular shape evolved into an elliptical one by the time the cavity reached the west solar limb, just prior to the eruption. It is important to note that the cavity centroid height slowly increased from 1.10 to 1.23 R$_S$ in association with a slow expansion in the cavity size as it rotated from the east to west solar limb (see table \ref{tab:table1}). \cite{gibson2006} demonstrated that an expanding flux rope may achieve an equilibrium configuration when the forces causing the expansion are counterbalanced by the confining magnetic tension forces exerted by the overlying magnetic field. As the magnetic field strength drops off radially with height, it is expected that the flux rope equilibrium will be governed by a lateral confinement rather than a vertical one, resulting in an elliptical morphology of the flux-rope cross-section \citep{Gibson2015}. Observations of the slowly rising and expanding cavity morphology presented in this work reveal that the cavity undergoes a series of stable equilibria during its quiescent phase. As the quiescent cavity expands and reaches higher heights in the corona it becomes more elliptical due to the excess magnetic pressure in the lateral direction exerted by the overlying magnetic field.

Interestingly, during the eruptive phase the elliptical cavity morphology again transformed back into a close to circular shape when the cavity centroid height crossed $\approx $ 4 R$_S$ in the FOV of LASCO C3. This implies that the confining magnetic tension forces fall off more rapidly with increasing radial height in comparison to the internal forces, causing the cavity expansion. Therefore, the domination of the internal expansion forces over the external magnetic tension force results in an isotropic expansion of the cavity which makes the cavity morphology close to circular in shape. This scenario is in agreement with the earlier findings of \citet{chen2000} where it has been shown that the role of magnetic tension forces becomes less significant compared to the drag and hoop force after the main acceleration phase of the CME which tends to occur below 2-3 R$_S$ \citep{chen2003,joshi2011}. 

The slowly rising and expanding phase of the quiescent cavity (see Table \ref{tab:table1}) during its passage from the east to west solar limb holds the intriguing clues to the underlying magnetostatic equilibria of the cavity system. During this stage, the injection of helical poloidal flux through flows or flux emergence from the lower boundary may gradually increase the toroidal current of the associated flux-rope which results in a gradual build-up in Lorentz self-force of the cavity system \citep{chen2006}. Therefore, the cavity centroid height showed a slow rise (Table \ref{tab:table1}) due to the gradual increase in the upward Lorentz self force against the downward magnetic tension force exerted by the overlying magnetic field. Examining the decay index-profiles during the quiescent phase of the cavity, we have found that the decay-index value ($<$1.0) at the cavity centroid height resided well below the critical value (1.2 to 1.5) for the onset of torus instability. This implies that although there was a slow build-up in the Lorentz self-force of the cavity system, it was not strong enough to overcome the downward magnetic tension forces due to the strong overlying magnetic-field strength. However, when the cavity appeared on the west solar limb, its centroid height attained a decay index value (1.3 $\pm$ .01) which belongs to the critical range (1.2 to 1.5) for the onset of a torus instability and eventually it erupted from the west solar limb. Therefore, we conclude that the decay-index value at the cavity centroid height can be used as a good indicator to examine whether a cavity will lead to an eruption or not.        

In what follows, we summarize the answers to the key questions (outlined in Section \ref{intro}) which have been addressed in this study:

(i) The slowly rising and expanding phase of the quiescent cavity hold important clues to its morphological evolution which reveals that the cavity undergoes a sequence of quasi-static equilibria during its long-lived quiescent phase. As the slowly rising cavity reaches higher heights in the corona, its morphology transforms from a nearly circular to an elliptical shape due to the excess magnetic pressure in the lateral direction exerted by the overlying magnetic field.

(ii) Throughout the quiescent phase, the cavity centroid height resided well below the critical limit for the onset of torus instability. However, the cavity centroid reached the critical height for the onset of torus instability just prior to the eruption. This critical height (1.23 R$_S$) in the lower corona determines the initiation height of the associated CME.

(iii) Evaluating the decay-index profiles at both the top-most part of the filament and the cavity centroid height, we conclude that  the cavity centroid should be used as the preferable location to evaluate the decay index value in order to reduce the discrepancy between the observational and theoretical critical limit for n$_{critical}$. Moreover, the decay index value of the cavity-centroid height can be used as a good indicator to determine whether the cavity will result in an eruption or not.

(iv) Using the combined FOV of SWAP, LASCO C2 and C3, we have shown how an EUV cavity evolved into a white-light cavity, as one of the three part structures of a CME, while maintaining the same spatial relationship with the underlying prominence material.

(v) The kinematic study of the cavity evolution successfully captures the two distinct phases of acceleration, where the impulsive acceleration phase was observed at 1.67 $\pm$ 0.08 R$_S$. The exponent value for the polynomial fit to the height-time profile of the erupting cavity reveals that the impulsive acceleration phase was most likely to be driven by the torus instability. However during the residual acceleration phase, the Lorentz self-force was less dominant and the kinematic evolution became strongly dependent on the drag force.

(vi) The significant non-radial motion observed in the SWAP FOV shows a strong deflection of the CME at lower coronal height ($\approx$ 1.3 R$_S$). A deflection ($\approx$ 40 $^\circ$ towards the equator) of the CME trajectory from higher to lower latitudes indicates that the polar coronal hole has significant influence on the CME kinematics, deflecting the CME towards the heliospheric current-sheet.
  
(vii) Finding the critical height above which the CME undergoes a self-similar expansion is one of the important results of this study. Interestingly, the critical height (2.2 $\pm$ 0.2 R$_S$) below which the cavity exhibited non-self similar expansion, points to the spatial scale of magnetic field lines fundamentally changing from a closed to an open regime, as they open into the solar wind.

These findings can be statistically validated with a larger dataset of events, which is beyond the scope of this current paper. In the future, we plan to carry out a statistical study on the erupting cavities in order to get deeper insight into the CME initiation mechanism.


We thank the referee for helpful comments that improved the quality of this manuscript. RS and NS would like to acknowledge the PROBA2 Guest Investigator program grant received to carry out this work. The PROBA2 Guest Investigator program grant and SWAP is a project of the Centre Spatial de Liege and the Royal Observatory of Belgium funded by the Belgian Federal Science Policy Office (BELSPO). E.D. acknowledges support from the Belgian Federal Science Policy Office (BELSPO) through the ESA-PRODEX programme, grant No. 4000120800. The authors also acknowledge the use of the data from SDO/AIA and STEREO/EUVI.\\
\bibliographystyle{yahapj}
\bibliography{sarkar2}  

 
\end{document}